\documentclass[a4paper,onecolumn,11pt,accepted=2023-02-06]{quantumarticle}
\pdfoutput=1
\usepackage{booktabs}
\usepackage{threeparttable}
\usepackage[utf8]{inputenc}
\usepackage[english]{babel}
\usepackage[T1]{fontenc}
\usepackage{amsmath}
\usepackage{hyperref}
\usepackage[numbers,sort&compress]{natbib}
\usepackage{tikz}
\usepackage{lipsum}

\usepackage{subcaption}
\usepackage{algorithmicx}
\usepackage{algorithm}
\usepackage{algpseudocode}
\usepackage{graphics}

\begin{document}

\title{Enabling Multi-programming Mechanism for Quantum Computing in the
	NISQ Era}

\author{Siyuan Niu}
\affiliation{LIRMM, University of Montpellier, 34095 Montpellier, France}
\orcid{0000-0003-4683-381X}
\email{siyuan.niu@lirmm.fr}
\author{Aida Todri-Sanial}
\affiliation{LIRMM, University of Montpellier, 34095 Montpellier, CNRS, France}
\affiliation{Eindhoven University of Technology, 5612 AE, Eindhoven, Netherlands}
\email{a.todri.sanial@tue.nl}
\orcid{0000-0001-8573-2910}

\maketitle

\begin{abstract}
   NISQ devices have several physical limitations and unavoidable noisy quantum operations, and only small circuits can be executed on a quantum machine to get reliable results. This leads to the quantum hardware under-utilization issue. Here, we address this problem and improve the quantum hardware throughput by proposing a Quantum Multi-programming Compiler (QuMC) to execute multiple quantum circuits on quantum hardware simultaneously. This approach can also reduce the total runtime of circuits. We first introduce a parallelism manager to select an appropriate number of circuits to be executed at the same time. Second, we present two different qubit partitioning algorithms to allocate reliable partitions to multiple circuits -- a greedy and a heuristic. Third, we use the Simultaneous Randomized Benchmarking protocol to characterize the crosstalk properties and consider them in the qubit partition process to avoid the crosstalk effect during simultaneous executions. Finally, we enhance the mapping transition algorithm to make circuits executable on hardware using a decreased number of inserted gates. We demonstrate the performance of our QuMC approach by executing circuits of different sizes on IBM quantum hardware simultaneously. We also investigate this method on VQE algorithm to reduce its overhead.
\end{abstract}

\section{Introduction}
\label{sec:introduction}

Quantum computing promises to achieve an exponential speedup to tackle certain computational tasks compared with the classical computers~\cite{doi:10.1137/S0097539795293172,lanyon2010towards,kerenidis2020quantum}. Quantum technologies are continuously improving, and IBM recently released the largest quantum chip with 127 qubits. But, current quantum devices are still qualified as Noisy Intermediate-Scale Quantum (NISQ) hardware~\cite{Preskill2018quantumcomputingin}, with several physical constraints. For example, for superconducting devices, which we target in this paper, connections are only allowed between two neighbouring qubits. Besides, the gate operations of NISQ devices are noisy and have unavoidable error rates. As we do not have enough number of qubits to realize Quantum Error Correction~\cite{calderbank1996good}, only small circuits with limited depth can obtain reliable results when executed on quantum hardware, which leads to a waste of hardware resources. 

With the growing demand to access quantum hardware, several companies such as IBM, Rigetti, and IonQ provide cloud quantum computing systems enabling users to execute their jobs on a quantum machine remotely. However, cloud quantum computing systems have some limitations. First, there exists a latency when submitting jobs. Second, there are a large number of jobs pending on the quantum device in general, so that users need to spend a long time waiting in the queue.

The low hardware usage and long waiting time lead to a timely issue: how do we use quantum hardware more efficiently while maintaining the circuit fidelity? As the increase of hardware qubit number and the improvement of qubit error rates, the multi-programming problem was introduced by~\cite{das2019case, liu2021qucloud} to address this issue. It has been demonstrated that the utilization (usage/throughput) of NISQ hardware can be enhanced by executing several circuits at the same time. However, their results showed that when executing multiple quantum circuits simultaneously, the activity of one circuit can negatively impact the fidelity of others, due to the difficulty of allocating reliable regions to each circuit, higher chance of crosstalk error, etc. Previous works~\cite{das2019case, liu2021qucloud} have left these issues largely unexplored and have not addressed the problem holistically such that the circuit fidelity reduction cannot be ignored when executing simultaneously. Moreover, detrimental crosstalk impact for multiple parallel instructions has been reported in ~\cite{murali2020software,ash2020analysis,ash2020experimental} by using Simultaneous Randomized Benchmarking (SRB)~\cite{gambetta2012characterization}. In the presence of crosstalk, gate error can be increased by an order of magnitude.  Ash-Saki et al.~\cite{ash2020analysis} even proposed a fault-attack model using crosstalk in a multi-programming environment. Therefore, crosstalk is considered in the multi-programming framework~\cite{ohkura2021simultaneous}. 

Multi-programming, if done in an ad-hoc way would be detrimental to fidelity, but if done carefully, it can be a very powerful technique to enable parallel execution for important quantum algorithms such as Variational Quantum Algorithms (VQAs)~\cite{cerezo2021variational}. For example, the multi-programming mechanism can enable to execute several ansatz states in parallel in one quantum processor, such as in Variational Quantum Eigensolver (VQE)~\cite{peruzzo2014variational,kandala2017hardware}, Variational Quantum Linear Solver (VQLS)~\cite{bravo2020variational}, or Variational Quantum Classifier (VQC)~\cite{havlivcek2019supervised} with reliability. It is also general enough to be applied to other quantum circuits regardless of applications or algorithms. More importantly, it can build the bridge between NISQ devices to large-scale fault-tolerant devices.

In this work, we address the problem of multi-programming by proposing a novel Quantum Multi-programming Compiler (QuMC), taking the impact of hardware topology, calibration data, and crosstalk into consideration. Our major contributions can be listed as follows:

\begin{itemize}
	\item We introduce a parallelism manager that can select the optimal number of circuits to execute simultaneously on the hardware without losing fidelity.
	
	\item We design two different qubit partition algorithms to allocate reliable partitions to multiple circuits. One is greedy, which provides the optimal choices. The other one is based on a heuristic that can give nearly optimal results and significantly reduce the time complexity.
	
	\item We consider crosstalk effect during the partition process to achieve crosstalk mitigation during simultaneous executions. This is the first crosstalk-aware partition algorithm.
	
	\item We improve the mapping transition step to execute multiple quantum circuits on quantum hardware with a reduced number of additional gates and better fidelity.
	
	\item  We present a use case of applying our multi-programming framework to the VQE algorithm to reduce its overhead, which demonstrates the application of multi-programming on NISQ algorithms.
	
\end{itemize}

We evaluate our algorithm on real quantum hardware by first executing circuits of different sizes at the same time, and then investigating it on VQE to estimate the ground state energy of deuteron. To the best of our knowledge, this is the first attempt to propose a complete multi-programming process flow for executing an optimal number of workloads in parallel ensuring the output fidelity by analyzing the hardware limitations, and the first demonstration of multi-programming application on NISQ algorithms.

\section{Background}
\subsection{NISQ computing}
Quantum computing has made huge progress in recent years. IBM launched the first cloud-based quantum computing service with a 5-qubit quantum machine in 2016, and the hardware qubit number reached 127 in only five years. In the meanwhile, the capabilities and error rates of the quantum hardware are continuously improving such that the Quantum Volume~\cite{cross2019validating} arrived 128 for IBM quantum machines. However, today's quantum computers are considered as NISQ devices yet. The hardware topology is limited and the qubits are prone to different errors, such as (1) coherent errors due to the fragile nature of qubits, (2) operational errors including gate errors and measurement errors (readout errors), (3) crosstalk errors that violate the isolated qubit state due to the parallel operations on other qubits. NISQ computing still promises to realize quantum advantages using variational quantum algorithms despite the errors. Cloud-based quantum computing services facilitate researchers and developers to work in this area. However, it causes some online traffic. For example, there are usually more than 100 jobs pending on IBM Q 27 Toronto, which requires several hours to retrieve the result. Therefore, efficient and reliable cloud quantum computing services are demanded while taking good care of hardware utilization and qubit errors.

\begin{figure}
	\centering
	\begin{subfigure}{0.4\columnwidth}
		\centering
		\includegraphics[scale=0.8]{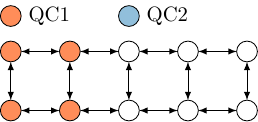}
		\label{fig:mul1}
		\caption{}
	\end{subfigure}
	\begin{subfigure}{0.4\columnwidth}
		\centering
		\includegraphics[scale=0.8]{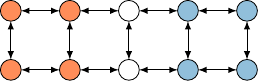}
		\label{fig:mul2}
		\caption{}
	\end{subfigure}
	
	\caption{An example of the multi-programming mechanism. 
		(a) A four-qubit circuit is executed on a 10-qubit device. The hardware throughput is 40\%.
		(b) Two four-qubit circuits are executed on the same device in parallel. The hardware throughput becomes 80\%.}
	
	\label{fig:mul}
	
\end{figure}

\subsection{Multi-programming mechanism}
The idea of the multi-programming mechanism is quite simple: executing several quantum circuits in parallel on the same quantum hardware. An example is shown in Fig.~\ref{fig:mul}. Note that, the simultaneous circuits can always be scheduled using As Late As Possible (ALAP) method, allowing qubits to remain in the ground state as long as possible to avoid additional decoherence error caused by circuits with different depths. Since the waiting time is usually much longer than the circuit execution time, the difference between execution time for circuits with different depths can be ignored (see experimental demonstration in Section~\ref{sec:depth}). By executing two circuits at the same time, the hardware throughput doubles and the total runtime (waiting time + execution time) is reduced twice. It is not trivial to achieve the multi-programming mechanism. The main concern is how to trade-off between the circuit output fidelity and the hardware throughput (also indicates the reduction of total runtime). Even though it is possible to simply combine several programs to one large circuit and compile it directly, it has been shown in~\cite{liu2021qucloud} that the circuit fidelity is decreased significantly due to the unfair allocation of partitions, unawareness of increased crosstalk, inflexibility of reverting back to independent executions for the case of serious fidelity drop, etc. Therefore, a new compilation technique for the multi-programming mechanism is required. Several problems need to be addressed to enable the multi-programming mechanism: (1) Find an appropriate number of circuits to be executed simultaneously such that the hardware throughput is improved without losing fidelity. (2) Allocate reliable partitions of the hardware to all the simultaneous circuits to make them execute with high fidelity. (3) Transform multiple circuits to make them executable on the hardware. (4) Reduce the interference between simultaneous circuit executions to lower the impact of crosstalk. 

\subsection{State of the art}
The multi-programming mechanism was first proposed in~\cite{das2019case} by developing a Fair and Reliable Partitioning (FRP) method. Liu et al. improved this mechanism and introduced QuCloud~\cite{liu2021qucloud}. There are some limitations for the two works: (1) Hardware topology and calibration data are not fully analyzed, such that allocation is sometimes done on unreliable or sparse-connected partitions ignoring the robust qubits and links. (2) These works use only \texttt{SWAP} gate for the mapping transition process and the modified circuits always have a large number of additional gates. (3) Crosstalk is not considered when allocating partitions for circuits. For example, the X-SWAP scheme~\cite{liu2021qucloud} can only be performed when circuits are allocated to neighbouring partitions, which is the case of more crosstalk. Ohkura et al. designed palloq~\cite{ohkura2021simultaneous}, a crosstalk detection protocol that reveals the crosstalk impact on multi-programming. A similar idea of Concurrent Quantum Circuit Sampling (CQCS) ~\cite{resch2021accelerating} was proposed to increase the hardware usage by executing multiple instances of the same program simultaneously. The concept of multi-programming was also explored in quantum annealers of DWAVE systems to solve several QUBO instances in parallel~\cite{pelofske2021parallel}. 

In our work, we focus on the multi-programming mechanism and propose QuMC framework with different crosstalk-aware partition methods and mapping transition algorithm to increase the hardware usage while maintaining the circuit fidelity.

\section{Our multi-programming framework}
Our proposed QuMC workflow is schematically shown in Fig.~\ref{fig:workflow}, which includes the following steps:

\begin{itemize}
	
	\begin{figure*}[!t]
		\centering
		\includegraphics[scale=0.5]{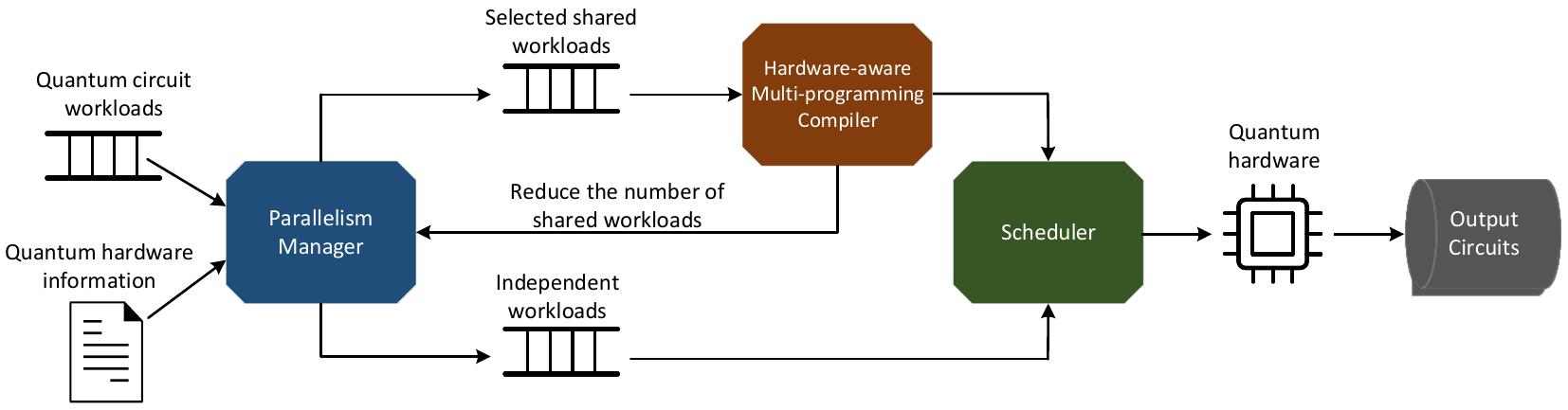}
		\caption{Overview of our proposed QuMC framework. The input layer includes the quantum hardware information and multiple quantum circuit workloads. The parallelism manager decides whether to execute circuits simultaneously or independently. For simultaneous executions, it works with the hardware-aware multi-programming compiler to select an optimal number of shared workloads to be executed in parallel. These circuits are allocated to reliable partitions and then passed to the scheduler. It makes all the circuits executable on the quantum hardware and we can obtain the results of the output circuits.
		}
		\label{fig:workflow}
	\end{figure*}
	
	\item Input layer. It contains a list of small quantum circuits written in OpenQASM language~\cite{cross2017open}, and the quantum hardware information, including the hardware topology, calibration data, and crosstalk effect.

	\item  Parallelism manager. It can determine whether executing circuits concurrently or separately.  If the simultaneous execution is allowed, it can further decide the number of circuits to be executed on the hardware at the same time without losing fidelity based on the fidelity metric included in the hardware-aware multi-programming compiler. 
	
	\item  Hardware-aware multi-programming compiler. Qubits are partitioned to several reliable regions and are allocated to different quantum circuits using qubit partition algorithms. Then, the partition fidelity is evaluated by the post qubit partition process. We introduce a fidelity metric here, which helps to decide whether this number of circuits can be executed simultaneously or the number needs to be reduced.
	
	\item Scheduler. The mapping transition algorithm is applied and circuits are transpiled to be executable on real quantum hardware. 
	
	\item  Output layer. Output circuits are executed on the quantum hardware simultaneously or independently according to the previous steps and the experimental results are obtained.
\end{itemize}
In this paper, we only focus on IBM quantum architecture. Our QuMC method can be generally adapted to quantum hardware with nearest-neighbor connectivity and also allows parallel operations if applied to different qubits.
\section{Parallelism manager}

\begin{figure*}[!t]
	\centering
	\includegraphics[scale=0.6]{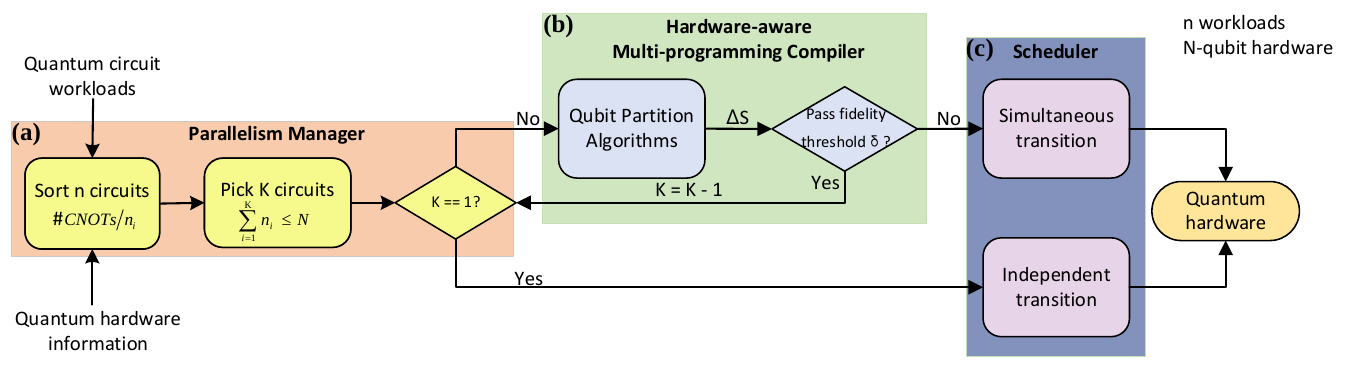}
	\caption{Process flow of each block that constitutes our QuMC approach. (a) The parallelism manager selects $K$ circuits according to their densities and passes them to the hardware-aware multi-programming compiler. (b) The qubit partition algorithms allocate reliable regions to multiple circuits. $\Delta S$ is the difference between partition scores when partitioning independently and simultaneously, which is the fidelity metric. $\delta$ is the threshold set by the user. The fidelity metric helps to select the optimal number of simultaneous circuits to be executed. (c) The scheduler performs mapping transition algorithm and makes quantum circuits executable on real quantum hardware.}	
	\label{fig:workflow2}
\end{figure*}		

In order to determine the optimal number of circuits that can be executed on the hardware in parallel without losing fidelity, here, we introduce the parallelism manager, shown in Fig.~\ref{fig:workflow2}(a).


Suppose we have a list of $n$ circuit workloads with $n_i$ qubits for each of them, that are expected to be executed on $N$-qubit hardware.
We define the circuit density metric as the number of \texttt{CNOTs} divided by the qubit number of the circuit, $\#CNOTs/n_i$, and the circuit with higher density is considered to be more subject to errors. Firstly, the circuits are ordered by their "density" metric. Note that, the users can also customize the order of circuits if certain circuits are preferred to have higher fidelities. Then, we pick $K$ circuits as the maximum number of circuits that can be executed on the hardware at the same time, $\sum_{n=1}^{K} n_i \leq N$. If $K$ is equal to one, then all the circuits should be executed independently. Otherwise, these circuits are passed to the hardware-aware multi-programming compiler. It works together with the parallelism manager to decide an optimal number of simultaneous circuits to be executed.

\section{Hardware-aware multi-programming compiler}

The hardware-aware multi-programming compiler contains two steps. First, perform qubit partitioning algorithm to allocate reliable partitions to multiple circuits. Second, compute the fidelity metric during post qubit partition process and work with parallelism manager to determine the number of simultaneous circuits. 
\subsection{Qubit partition}
We develop two qubit partition algorithms by accounting for the crosstalk, hardware topology, and calibration data. In this section, we first introduce a motivational example for qubit partition. Second, we explain the approach for crosstalk characterization. Finally, we present two qubit partition algorithms, one greedy and one heuristic. 
\subsubsection{Motivational example}

\begin{figure}
	\centering
	\begin{subfigure}{0.45\columnwidth}
		\centering
		\includegraphics[scale=0.7]{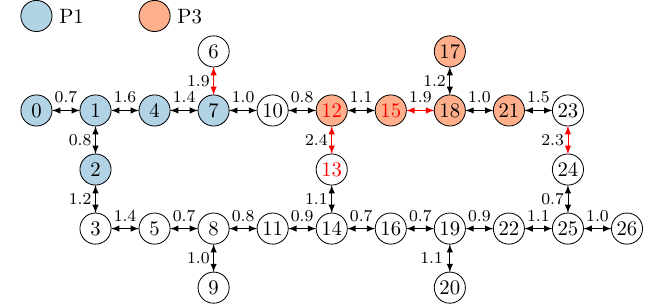}
		\label{fig:multi_1}
		\caption{}
	\end{subfigure}
	\hfil
	\begin{subfigure}{0.45\columnwidth}
		\centering
		\includegraphics[scale=0.7]{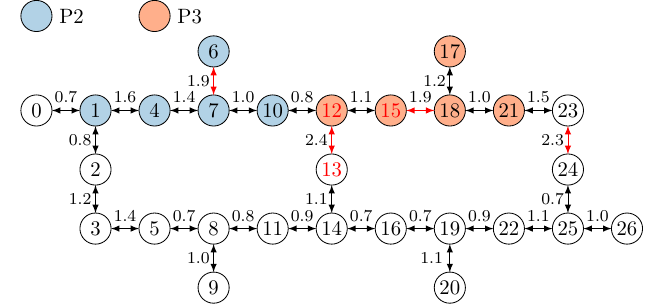}
		\label{fig:multi_2}
		\caption{}
	\end{subfigure}
	
	\caption{A motivational example of qubit partition problem. 
		(a) No crosstalk between partition P1 and partition P3.
		(b) Crosstalk exists between partition P2 and partition P3.
	}
	\label{fig:motivation}
	
\end{figure}

\begin{figure}
	
	\centering
	\includegraphics[scale=0.4]{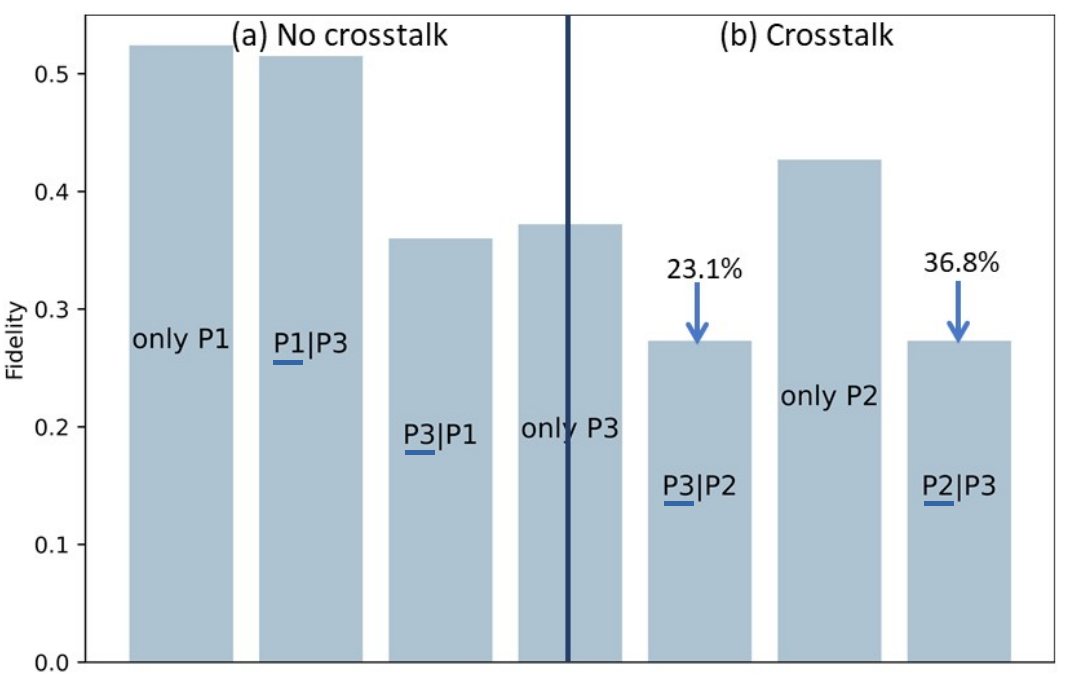}
	\label{fig:motivation_fidelity}
	
	\caption{Results of the motivational example. (a) No crosstalk corresponds to Fig.~\ref{fig:motivation}(a) where no crosstalk exists between P1 and P3. (b) Crosstalk corresponds to Fig.~\ref{fig:motivation}(b) where crosstalk exists between P2 and P3. Note that "only P1" means the fidelity of the circuit when it is executed independently on P1, whereas "P1|P3" means the fidelity of circuit on P1 when two circuits are executed on P1 and P3 simultaneously. }
	
	\label{fig:motivation_result}
\end{figure}

We consider two constraints when executing multiple circuits concurrently. First, each circuit should be allocated to a partition containing reliable physical qubits. Allocated physical qubits (qubits used in hardware) can not be shared among quantum circuits. Second, qubits can be moved only inside of their circuit partition during the routing process, in other words, qubits can be swapped within the same partition only. Note that, in this work, we performed routing inside of the reliable partition but other approaches can be applied as well such as to route to other neighboring qubits that are outside of the reliable partition.

Finding reliable partitions for multiple circuits is an important step in the multi-programming problem. In order to illustrate the impact of partitions with different error sources on the output fidelity, first, we execute a small circuit \texttt{alu-v0\_27} (the information of this circuit can be found in Table~\ref{fig:benchmarks}) on three different partitions independently to show the impact of operational error (including \texttt{CNOT} error and readout error): (1) Partition P1 with reliable qubits and links. (2) Partition P2 with unreliable links. (3) Partition P3 with unreliable links and qubits with high readout error rate. Note that, the \texttt{CNOT} error rate of each link is shown in Fig.~\ref{fig:motivation} and the unreliable links with high \texttt{CNOT} error rates and qubits with high readout error rates are highlighted in red. Second, we execute two of the same circuits simultaneously to show the crosstalk effect: (1) P1 and P3 without crosstalk (Fig.~\ref{fig:motivation}(a)). (2) P2 and P3 with crosstalk (Fig.~\ref{fig:motivation}(b)). For the sake of fairness, each partition has the same topology. It is important to note that if we have different topologies, the circuit output fidelity will also be different since the number of additional gates is strongly related to the hardware topology.

The result of the motivational example is shown in Fig.~\ref{fig:motivation_result}. The fidelity is calculated using PST metric explained in Section~\ref{metrics} and higher is better. For independent execution, we have P1 > P2 > P3 in terms of fidelity, which shows the influence of operational error on output fidelity. For simultaneous execution, the circuit fidelities are approximately the same for the two partitions P1 and P3 compared with the independent execution in the case of no crosstalk. Whereas, the fidelities are decreased by 36.8\% and 23.1\% respectively for P2 and P3 when the two circuits are executed simultaneously due to the crosstalk. This example demonstrates the importance of considering crosstalk effect in the multi-programming mechanism.

\subsubsection{Crosstalk effect characterization.}

Crosstalk is one of the major noise sources in NISQ devices, which can corrupt a quantum state due to quantum operations on other qubits~\cite{sarovar2020detecting}. There are two types of crosstalk. The first one is quantum crosstalk, which is caused by the always-on-ZZ interaction~\cite{mundada2019suppression,zhao2020high}. The second one is classical crosstalk caused by the incorrect control of the qubits. The calibration data provided by IBM do not include the crosstalk error. To consider the crosstalk effect in partition algorithms, we must first characterize it in the hardware. There are several protocols presented in~\cite{gambetta2012characterization,bialczak2010quantum,proctor2019direct,erhard2019characterizing} to benchmark the crosstalk effect in quantum devices. In this paper, we choose the mostly used protocol --  Simultaneous Randomized Benchmarking (SRB)~\cite{gambetta2012characterization} to detect and quantify the crosstalk between \texttt{CNOT} pairs when executing them in parallel.

\begin{figure}[!h]
	\centering
	\begin{subfigure}{0.45\columnwidth}
		\centering
		\includegraphics[scale=0.7]{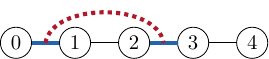} 
		\label{fig:one_hop}
		\caption{}
	\end{subfigure}
	\hfil
	\begin{subfigure}{0.45\columnwidth}
		\centering
		\includegraphics[scale=0.7]{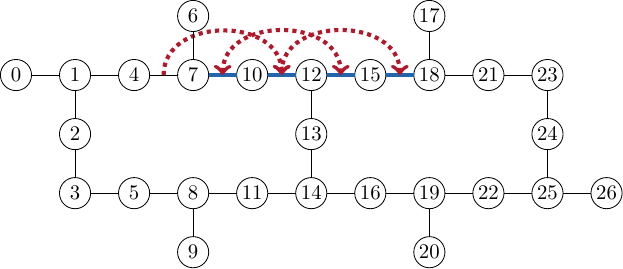}
		\label{fig:crosstalk}
		\caption{}
	\end{subfigure}
	
	\caption{Characterization of crosstalk effect. (a)  Crosstalk pairs separated by one-hop distance. The crosstalk pairs should be able to be executed at the same time. Therefore, they cannot share the same qubit. One-hop is the minimum distance between crosstalk pairs. (b) Crosstalk effect results of IBM Q 27 Toronto using SRB.  The arrow of the red dash line points to the \texttt{CNOT} pair that is affected significantly by crosstalk effect, e.g., $CX_{7,10}$ and $CX_{12,15}$ affect each other when they are executed simultaneously. In our experiments, $E(CX_{10,12}|CX_{4,7}) > 3 \times E(CX_{10,12})$, whereas $E(CX_{4,7}|CX_{10,12}) \approx 1.5 \times E(CX_{4,7})$. As we choose 3 as the factor to pick up pairs with strong crosstalk effect, there is no arrow at pair $CX_{4,7}$.}
	\label{fig:characterization}
\end{figure}

We characterize the crosstalk effect followed by the optimization methods presented in~\cite{murali2020software}. On IBM quantum devices, the crosstalk effect is significant only at one hop distance between \texttt{CNOT} pairs~\cite{murali2020software}, such as ($CX_{0,1} | CX_{2,3}$) shown in Fig.~\ref{fig:characterization}(a), when the control pulse of one qubit propagates an unwanted drive to the nearby qubits that have similar resonate frequencies. Therefore, we perform SRB only on \texttt{CNOT} pairs that are separated by one-hop distance. For those pairs whose distance is greater than one hop, the crosstalk effects are very weak and we ignore them. It allows us to parallelize SRB experiments of multiple \texttt{CNOT} pairs when they are separated by two or more hops. For example, in IBM Q 27 Toronto, the pairs ($CX_{0,1} | CX_{4,7}$), ($CX_{12,15} | CX_{17,18}$), ($CX_{5,8} | CX_{11,14}$) can be characterized in parallel. 

Previous works~\cite{murali2020software,ash2020analysis,9516713} show that, although the absolute gate errors vary every day, the pairs that have strong crosstalk effect remain the same across days. We confirm that validation by performing the crosstalk characterization on IBM Q 27 Toronto twice and we observe the similar behavior. The SRB experiment on \texttt{CNOT} pairs ($g_i|g_j$) gives  error rate $E(g_i|g_j)$ and $E(g_j|g_i)$. Here, $E(g_i|g_j)$ represents the correlated \texttt{CNOT} error rate of $g_i$ when $g_i$ and $g_j$ are executed in parallel. If there is a crosstalk effect between the two pairs, it will lead to $E(g_i|g_j) > E(g_i)$ or $E(g_j|g_i) > E(g_j)$. The crosstalk effect characterization is expensive and time costly. Some of the pairs do not have crosstalk effect whereas the correlated \texttt{CNOT} error affected the most by crosstalk effect is increased by more than five times. Therefore, we extract the pairs with significant crosstalk effect, i.e., $E(g_i|g_j) > 3 \times E(g_i)$ and only characterize these pairs when crosstalk properties are needed. We choose the same factor 3 to quantify the pairs with strong crosstalk error like~\cite{murali2020software}. The result of crosstalk effect characterization on IBM Q 27 Toronto is shown in Fig.~\ref{fig:characterization}(b).	

\subsubsection{Greedy sub-graph partition algorithm.}

We develop a Greedy Sub-graph Partition algorithm (GSP) for qubit partition process which is able to provide the optimal partitions for different quantum circuits.
The first step of the GSP algorithm is to traverse the overall hardware to find all the possible partitions for a given circuit. For example, suppose we have a five-qubit circuit, we find all the subgraphs of the hardware topology (also called coupling graph) containing five qubits as the partition candidates. Each candidate has a score to represent its fidelity depending on the topology and calibration data. The partition with the best fidelity is selected and all the qubits inside of the partition are marked as used qubits so they cannot be assigned to other circuits. For the next circuit, a subgraph with the required number of qubits is assigned and we check if there is an overlap on this partition to partitions of previous circuits. If not, the subgraph is a partition candidate for the given circuit and the same process is applied to each subsequent circuit. To account for crosstalk, we check if any pairs in a subgraph have strong crosstalk effect caused by the allocated partitions of other circuits. If so, the score of the subgraph is adjusted to take crosstalk error into account. 

In order to evaluate the reliability of a partition, three factors need to be considered: partition topology, error rates of two-qubit links, and readout error of each qubit. One-qubit gates are ignored for simplicity and because of their relatively low error rates compared to the other quantum operations. If there is a qubit pair in a partition that has strong crosstalk affected by other partitions, the \texttt{CNOT} error of this pair is replaced by the correlated \texttt{CNOT} error which takes crosstalk into account. Note that the most recent calibration data should be retrieved through the IBM Quantum Experience before each usage to ensure that the algorithm has access to the most accurate and up-to-date information. 
To evaluate the partition topology, we determine the longest shortest path (also called graph diameter) of the partition, denoted $L$. The smaller the longest shortest path is, the better the partition is connected. Eventually, fewer additional gates would be needed to connect two qubits in a well-connected partition.

\begin{algorithm}[H]
	\begin{algorithmic}[1]
		
		\caption{GSP algorithm} \label{algo1}
		\Require {Quantum circuit $QC$ 
			, Coupling graph $G$, Calibration data $C$,
			Crosstalk properties crosstalk\_props,
			Used\_qubits $q_{\text{used}}$ }
		\Ensure{A list of candidate partitions sub\_graph\_list}
		
		\State qubit\_num $\leftarrow$ $QC$.qubit\_num
		
		\State	Set sub\_graph\_list to empty list
		\For {sub\_graph $\in$ combinations($G$, qubit\_num)}
		\If{sub\_graph is connected}
			\If {$q_{\text{used}}$ is empty}
			\State	sub\_graph.Set\_Partition\_Score($G$, $C$, $QC$)
			\State	sub\_graph\_list.append(sub\_graph)
			
			\EndIf
			
			\If {no qubit in sub\_graph is in $q_{\text{used}}$}
			\State	crosstalk\_pairs $\leftarrow$ Find\_Crosstalk\_pairs(sub\_graph,\\\hspace{2.5cm} crosstalk\_props, $q_{\text{used}}$)             
			\State	sub\_graph.Set\_Partition\_Score($G$, $C$, $QC$, crosstalk\_pairs)
			\State	sub\_graph\_list.append(sub\_graph)
			
			\EndIf
			\EndIf
			\EndFor

			\State	\Return sub\_graph\_list 
			
		\end{algorithmic}
		
	\end{algorithm}

	We devise a fidelity score metric for a partition that is the sum of the graph diameter $L$, average \texttt{CNOT} error rate of the links times the number of \texttt{CNOTs} of the circuit, and the sum of the readout error rate of each qubit in a partition (shown in~\eqref{eq:1}). Note that the \texttt{CNOT} error rate includes the crosstalk effect if it exists.
	
	\begin{equation}
		Score_g = L + Avg_{CNOT} \times \#CNOTs + \sum_{Q_i \in P}R_{Q_i}
		\label{eq:1}
	\end{equation}
	
	The graph diameter $L$ is always prioritized in this equation, since it is more than one order of magnitude larger than the other two factors. The partition with the smallest fidelity score is selected. It is supposed to have the best connectivity and the lowest error rate. Moreover, the partition algorithm prioritizes the quantum circuit with a large density because the input circuits are ordered by their densities during the parallelism manager process. The partition algorithm is then called for each circuit in order. However, GSP algorithm is expensive and time costly. For small circuits, the GSP algorithm gives the best choice of partition. It is also useful to use it as a baseline to compare with other partition algorithms. For beyond NISQ, a better approach should be explored to overcome the complexity overhead.

	\subsubsection{Qubit fidelity degree-based heuristic sub-graph partition algorithm.}
	
	In order to reduce the overhead of GSP, we propose a Qubit fidelity degree-based Heuristic Sub-graph Partition algorithm (QHSP). It performs as well as GSP but without the large runtime overhead.

	In QHSP, when allocating partitions, we favor qubits with high fidelity. We define the fidelity degree of a qubit based on the \texttt{CNOT} and readout fidelities of this qubit as in~\eqref{eq:2}. 
	\begin{equation}
		F\_Degree_{Q_i} = \sum_{Q_j \in N(Q_i)} \lambda \times (1 - E(Q_i,Q_j) + (1 - R_{Q_i})
		\label{eq:2}
	\end{equation}
	$Q_j$ are the neighbour qubits connected to $Q_i$, $E$ is the \texttt{CNOT} error matrix which is constructed by applying the Floyd-Warshall algorithm to the hardware coupling graph with \texttt{CNOT} error rate as edge weights, and $R$ is the readout error rate. $\lambda$ is a user defined parameter to weight between the \texttt{CNOT} error rate and readout error rate. Such parameter is useful for two reasons: (1) Typically, in a quantum circuit, the number of \texttt{CNOT} operations is different from the number of measurement operations. Hence, the user can decide $\lambda$ based on the relative number of operations. (2) For some qubits, the readout error rate is one or more orders of magnitude larger than the \texttt{CNOT} error rate. Thus, it is reasonable to add a weight parameter.
	
	The fidelity degree metric reveals two aspects of a qubit. The first one is the connectivity of the qubit. The more neighbours a qubit has, the larger its fidelity degree is. The second one is the reliability of the qubit accounting \texttt{CNOT} and readout error rates.  Thus, the metric allows us to select a reliable qubit with good connectivity. Instead of trying all the possible subgraph combinations (as in the GSP algorithm), we propose a QHSP algorithm to build partitions that contain qubits with high fidelity degree while significantly reducing runtime.

	To further improve the algorithm, we construct a list of qubits with good connectivity as starting points. We sort all physical qubits by their physical node degree, which is defined as the number of links in a physical qubit. Note that, the physical node degree is different from the fidelity degree. Similarly, we also obtain the largest logical node degree of the logical qubit (qubits used in the quantum circuit) by checking the number of different qubits that are connected to a qubit through \texttt{CNOT} operations. Next, we compare these two metrics.
	
	Suppose the largest physical node degree is less than the largest logical node degree. In that case, it means that we cannot find a suitable physical qubit to map the logical qubit with the largest logical node degree that satisfies all the connections. In this case, we only collect the physical qubits with the largest physical node degree. Otherwise, the physical qubits whose physical node degree is greater than or equal to the largest logical node degree are collected as starting points. By limiting the starting points, this heuristic partition algorithm becomes even faster.

	\begin{algorithm}[H]%
		\begin{algorithmic}[1]%
			\caption{QHSP algorithm} \label{algo2}
			
			\Require{Quantum circuit $QC$ 
				, Coupling graph $G$, Calibration data $C$,
				Crosstalk properties crosstalk\_props,
				Used\_qubits $q_{\text{used}}$,
				Starting points starting\_points}
			\Ensure{A list of candidate partitions sub\_graph\_list}
			
			\State circ\_qubit\_num $\leftarrow$  $QC$.qubit\_num
			\State 	Set sub\_graph\_list to empty list
			\For{i $\in$ starting\_points}
			\State 		Set sub\_graph to empty list
			\State 		qubit\_num $\leftarrow$ $0$
			\While{qubit\_num $<$ circ\_qubit\_num}
			\If{sub\_graph is empty}
			\State sub\_graph.append(i)
			\State qubit\_num $\leftarrow$ qubit\_num + 1 
			\State continue
			\EndIf
			\State best\_qubit $\leftarrow$ find\_best\_qubit(sub\_graph, $G$, $C$)
			\If{best\_qubit $\neq$ None}
			\State sub\_graph.append(best\_qubit)
			\State qubit\_num $\leftarrow$ qubit\_num + 1 
			\State continue
			\EndIf
			\EndWhile
			\If{len(sub\_graph) = circ\_qubit\_num}
			\If{$q_{\text{used}}$ is empty}
			\State sub\_graph.Set\_Partition\_Score($G$,  $C$,
			$QC$)
			\State sub\_graph\_list.append(sub\_graph)
			\EndIf
			\If{no qubit in sub\_graph is in $q_{\text{used}}$}
			\State crosstalk\_pairs $\leftarrow$ Find\_Crosstalk\_pairs(sub\_graph,\\\hspace{2.5cm} crosstalk\_props, $q_{\text{used}}$)            
			\State sub\_graph.Set\_Partition\_Score($G$, $C$,  $QC$, crosstalk\_pairs)
			\State sub\_graph\_list.append(sub\_graph)
			\EndIf
			\EndIf
			
			\EndFor	
			
			\State \Return sub\_graph\_list 
			
		\end{algorithmic}
	\end{algorithm}
	
	For each qubit in the starting points list, the algorithm explores its neighbours and finds the neighbour qubit with the highest fidelity degree calculated in \eqref{eq:2}, and merges it into the sub-partition. Then, the qubit inside of the sub-partition with the highest fidelity degree explores its neighbour qubits and merges the best one. The process is repeated until the number of qubits inside of the sub-partition is equal to the number of qubits needed. This sub-partition is considered as a subgraph and is added to the partition candidates.

	After obtaining all the partition candidates,  we compute the fidelity score for each of them. As we start from a qubit with a high physical node degree and merge to neighbour qubits with a high fidelity degree, the constructed partition is supposed to be well-connected, hence, we do not need to check the connectivity of the partition using the longest shortest path $L$ as in \eqref{eq:1}, GSP algorithm. We can only compare the error rates. The fidelity score metric is simplified by only calculating the \texttt{CNOT} and readout error rates as in \eqref{eq:3} (crosstalk is included if it exists). It is calculated for each partition candidate and the best one is selected. 
	
	\begin{equation}
		Score_h = Avg_{CNOT} \times \#CNOTs + \sum_{Q_i \in P}R_{Q_i}
		\label{eq:3}
	\end{equation}

	\begin{figure}[!htp]
		\centering
		\begin{subfigure}{0.4\columnwidth}
			\centering
			\includegraphics[scale=0.85]{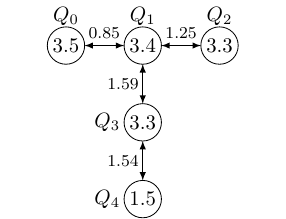}
			\label{fig:ibmq_valencia_info}
			\caption{}
		\end{subfigure}
		\hfil
		\begin{subfigure}{0.4\columnwidth}
			\centering
			\includegraphics[scale=0.8]{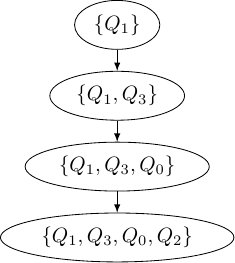}
			\label{fig:Qubit partition}
			\caption{}
		\end{subfigure}
		
		\caption{Example of qubit partition on IBM Q 5 Valencia for a four-qubit circuit using QHSP. Suppose the largest logical node degree of the target circuit is three. (a) The topology and calibration data of IBM Q 5 Valencia. The value inside of the node represents the readout error rate (in\%), and the value above the link represents the \texttt{CNOT} error rate (in\%). (b) Process of constructing a partition candidate using QHSP.}
		\label{fig:partition_example}
	\end{figure}
	
	\begin{table}[!htp]
		\caption{\label{fig:qubit degree}The physical node degree and the fidelity degree of each qubit on IBM Q 5 Valencia.} 
		\centering
		\begin{tabular}{@{}*{6}{l}}
			\toprule                               
			Qubit & $Q_0$ & $Q_1$ & $Q_2$ &  $Q_3$ & $Q_4$\cr
			\midrule
			Fidelity degree & $1.96$ & $3.93$ & $1.95$ & $2.94$ & $1.97$\cr
			
			Physical node degree & 1 & 3 & 1 & 2 & 1 \cr
			
			\bottomrule
		\end{tabular}
		
	\end{table}
	
	Fig.~\ref{fig:partition_example} shows an example of applying QHSP on IBM Q 5 Valencia (ibmq\_valencia) for a four-qubit circuit. The calibration data of IBM Q 5 Valencia, including readout error rate and \texttt{CNOT} error rate are shown in Fig.~\ref{fig:partition_example}(a). We set $\lambda$ to two and the physical node degree and the fidelity degree of qubit calculated by \eqref{eq:2} are shown in Table~\ref{fig:qubit degree}. Suppose the largest logical node degree is three. Therefore, $Q_1$ is selected as the starting point since it is the only physical qubit that has the same physical node degree as the largest logical node degree. It has three neighbour qubits: $Q_0$, $Q_2$, and $Q_3$. $Q_3$ is merged into the sub-partition because it has the highest fidelity degree among neighbour qubits. The sub-partition becomes $\{Q_1, Q_3\}$. As the fidelity degree of $Q_1$ is larger than $Q_3$, the algorithm will again select the left neighbour qubit with the largest fidelity degree of $Q_1$, which is $Q_0$. The sub-partition becomes $\{Q_1, Q_3, Q_0\}$. $Q_1$ is still the qubit with the largest fidelity degree in the current sub-partition, its neighbour qubit -- $Q_2$ is merged. The final sub-partition is $\{Q_1,Q_3,Q_0,Q_2\}$ and it can be considered as a partition candidate. The merging process is shown in Fig.~\ref{fig:partition_example}(b).
	
	\subsubsection{Runtime analysis}
	
	Let $n$ be the number of hardware qubits (physical qubits) and $k$ the number of circuit qubits (logical qubits) to be allocated a partition. The GSP algorithm selects all the combinations of $k$ subgraphs from $n$-qubit hardware and takes $O(C(n,k))$ time, which is $O(n \, choose \, k)$. For each subgraph, it computes its fidelity score including calculating the longest shortest path, which scales at $O(k^3)$. It ends up being equivalent to $O(k^3min(n^k, n^{n-k}))$. In most cases, the number of circuit qubits is less than the number of hardware qubits, thus the time complexity becomes $O(k^3n^k)$. It increases exponentially as the number of circuit qubits augments.
	
	The QHSP algorithm starts by collecting a list of $m$ starting points where $m \leq n$. To get the starting points, we sort the $n$ physical qubits by their physical node degree, which takes $O(nlog(n))$. Then, we iterate over all the gates of the circuit (e.g., circuit has $g$ gates) and sort the $k$ logical qubits according to the logical node degree, which takes $O(g+klog(k))$. Next, for each starting point, it iteratively merges the best neighbour qubit until each sub-partition contains $k$ qubits. To find the best neighbour qubit, the algorithm finds the best qubit in a sub-partition and traverses all its neighbours to select the one with the highest fidelity degree. Finding the best qubit in the sub-partition is $O(p)$ where $p$ is the number of qubits in a sub-partition. The average number of qubits $p$ is $k/2$, so this process takes $O(k)$ time on average. Finding the best neighbour qubit is $O(1)$ because of the nearest-neighbor connectivity of superconducting devices. Overall, the QHSP takes $O(mk^2 + nlog(n) + g + klog(k))$ time, and it can be truncated to $O(mk^2 + nlog(n) + g)$, which is polynomial. 
	
	\subsection{Post qubit partition}
	
	By default the multi-programming mechanism reduces circuit fidelity compared to standalone circuit execution mode. If the fidelity reduction is significant, circuits should be executed independently or the number of simultaneous circuits should be reduced even though the hardware throughput can be decreased as well. Therefore, we consistently check the circuit fidelity difference between independent versus concurrent execution.
	
	We start with the qubit partition process for each circuit independently and obtain the fidelity score of the partition. Next, this qubit partition process is applied to these circuits to compute the fidelity score when executing them simultaneously. The difference between the fidelity scores is denoted $\Delta S$, which is the fidelity metric. If $\Delta S$ is less than a specific threshold $\delta$,  it means simultaneous circuit execution does not significantly detriment the fidelity score, thus circuits can be executed concurrently, otherwise, independently or reduce the number of simultaneous circuits. The fidelity metric and the parallelism manager help determine the optimal number of simultaneous circuits to be executed.
	
	\section{Scheduler}
	The scheduler includes the mapping algorithm to make circuits executable on real quantum hardware. 
	\subsection{Mapping transition algorithm}
	
	Two steps are needed to make circuits hardware-compliant: initial mapping and mapping transition. The initial mapping of each circuit is created while taking into account swap error rate and swap distance, and the initial mapping of the simultaneous mapping transition process is obtained by merging the initial mapping of each circuit according to its partition. We improve the mapping transition algorithm proposed in \cite{niu2020hardware} by modifying the heuristic cost function to better select the inserted gate. We also introduce the \texttt{Bridge} gate to the simultaneous mapping transition process for multi-programming.

	First, each quantum circuit is transformed into a more convenient format -- Directed Acyclic Graph (DAG) circuit, which represents the operation dependencies of the circuit without considering the connectivity constraints. Then, the compiler traverses the DAG circuit and goes through each quantum gate sequentially. The gate that does not depend on other gates (i.e., all the gates before execution) is allocated to the first layer, denoted $F$. The compiler checks if the gates on the first layer are hardware-compliant. The hardware-compliant gates can be executed on the hardware directly without modification. They are added to the scheduler, removed from the first layer and marked as executed. If the first layer is not empty, which means some gates are non-executable on hardware, a \texttt{SWAP} or \texttt{Bridge} gate is needed. We collect all the possible \texttt{SWAPs} and \texttt{Bridges}, and use the cost function $H$ (see \eqref{eq:5}) to find the best candidate. The process is repeated until all the gates are marked as executed.
	
	A \texttt{SWAP} gate requires three \texttt{CNOTs} and inserting a \texttt{SWAP} gate can change the current mapping. Whereas a \texttt{Bridge} gate requires four \texttt{CNOTs} and inserting a \texttt{Bridge} gate does not change the current mapping. It can only be used to execute a \texttt{CNOT} when the distance between the control and the target qubits is exactly two. Both gates need three supplementary \texttt{CNOTs}. A \texttt{SWAP} gate is preferred when it has a positive impact on the following gates, allocated in the extended layer $E$, i.e., it makes these gates executable or reduces the distance between control and target qubits. Otherwise, a \texttt{Bridge} gate is preferred. 
	
	A cost function $H$ is introduced to evaluate the cost of inserting a \texttt{SWAP} or \texttt{Bridge}. We use the following distance matrix (see \eqref{eq:4}) as in~\cite{niu2020hardware} to quantify the impact of the \texttt{SWAP} or \texttt{Bridge} gate,
	
	\begin{equation}
		D = \alpha_1 \times S + \alpha_2 \times \mathcal{E}
		\label{eq:4}
	\end{equation}
	where $S$ is the swap distance matrix and $\mathcal{E}$ is the swap error matrix. We set $\alpha_1$ and $\alpha_2$ to 0.5 to equally consider the swap distance and swap error rate. In~\cite{niu2020hardware}, only the impact of a \texttt{SWAP} and \texttt{Bridge} on other gates (first and extended layer) was considered without considering their impact on the gate itself. As each of them is composed of either three or four \texttt{CNOTs}, their impact cannot be ignored. Hence, in our simultaneous mapping transition algorithm, we take self impact into account and create a list of both \texttt{SWAP} and \texttt{Bridge} candidates, labeled as "tentative gates". The heuristic cost function is as:
	
	
	\begin{equation}
		H = \frac{1}{\vert F + N_{Tent} \vert}(\sum_{g \in F} D[\pi(g.q_1)][\pi(g.q_2)]     + \sum_{g \in Tent}D[\pi(g.q_1)][\pi(g.q_2)]) 
		+ W \times \frac{1}{\vert E \vert}\sum_{g \in E}D[\pi(g.q_1)][\pi(g.q_2)]
		\label{eq:5}
	\end{equation}

	where $W$ is the parameter that weights the impact of the extended layer, $N_{Tent}$ is the number of gates of the tentative gate, $Tent$ represents a \texttt{SWAP} or \texttt{Bridge} gate, and $\pi$ represents the mapping. \texttt{SWAP} gate has three \texttt{CNOTs}, thus $N_{Tent}$ is three and we consider the impact of three \texttt{CNOTs} on the first layer. The mapping is the new mapping after inserting a \texttt{SWAP}. For \texttt{Bridge} gate, $N_{Tent}$ is four and we consider four \texttt{CNOTs} on the first layer, and the mapping is the current mapping as \texttt{Bridge} gate does not change the current mapping. We weight the impact on the extended layer to prioritize the first layer. This cost function can help the compiler select the best gate to insert between a \texttt{SWAP} and \texttt{Bridge} gate.
	
	Our simultaneous mapping transition algorithm outperforms HA~\cite{niu2020hardware} thanks to the modifications of the cost function while not changing its asymptotic complexity. Let $n$ be the number of hardware qubits, $g$ the \texttt{CNOT} gates in the circuit. The simultaneous mapping transition algorithm takes $O(gn^{2.5})$ assuming nearest-neighbor chip connectivity and an extended layer $E$ with at most $O(n)$ \texttt{CNOT} gates. The detailed explanation about the complexity can be found in~\cite{niu2020hardware}.

	\begin{algorithm}[H]
		\begin{algorithmic}[1]%
			\caption{Simultaneous mapping transition algorithm} \label{algo3}
			
			\Require{Circuits $DAGs$ 
				, Coupling graph $G$, Distance matrices $Ds$, Initial mapping $\pi_i$, First layers $Fs$
			}
			\Ensure{Final schedule}
			
			\State $\pi_c$ $\leftarrow$ $\pi_i$ 
			\While{ not all gates are executed}
			\State Set swap\_bridge\_lists to empty list
			\For{$F_i$ in $Fs$}
			\For{gate in $F_i$}
			\If{gate is hardware-compliant}
			\State schedule.append(gate)
			\State Remove gate from $F_i$
			\EndIf
			\EndFor
			\If{$F_i$ is not empty}
			\State 
			swap\_bridge\_candidate\_list  $\leftarrow$ FindSwapBridgePairs($F_i$, $G$)
			\State
			swap\_bridge\_lists.append(swap\_bridge\_candidate\_list)
			\EndIf
			\EndFor
			\For{swap\_bridge\_candidate\_list $\in$ swap\_bridge\_lists}
			\For{$g_\text{{tmp}}$ $\in$ swap\_bridge\_candidate\_list}
			\State $\pi_\text{{tmp}}$ $\leftarrow$ Map\_Update($g_\text{{tmp}}$, $\pi_c$)
			\State $H_\text{{basic}}$ $\leftarrow$ $0$
			\For{gate $\in$ $F_i$}
			\State $H_\text{{basic}}$ $\leftarrow$ $H_\text{{basic}}$ + $D_i$(gate, $\pi_\text{{tmp}}$)
			\EndFor
			\State $H_\text{{tentative}}$ $\leftarrow$ $g_\text{{tmp}}$.$cost$($G$, $D_i$,  $\pi_\text{{tmp}}$)
			
			\State Update the extended layer $E$
			\State $H_\text{{extend}}$ $\leftarrow$ $0$
			\For{gate $\in$ $E$}
			\State $H_\text{{extend}}$ $\leftarrow$ $H_\text{{extend}}$ + $D_i$(gate, $\pi_\text{{tmp}}$)
			\EndFor	
			
			\State 
			$H$ $\leftarrow \frac{1}{\vert F + H_\text{{tentative}} \vert} (H_\text{{basic}} + H_\text{{tentative}}) + \frac{W}{\vert E \vert} H_\text{{extend}}$
			
			\EndFor
			\State Choose the best gate $g_n$ according to $H$
			\State $\pi_c$ $\leftarrow$ Map\_Update($g_n$, $\pi_c$)
			\EndFor
			\State Update $Fs$
			\EndWhile

			\State	\Return schedule		
		\end{algorithmic}
		
	\end{algorithm}
	
	\section{Evaluation}
	
	In this section, we compare our QuMC method with the state of the art and showcase its different applications.
	\subsection{Methodology}
	\subsubsection{Metrics}
	\label{metrics}
	Here are the explanations of the metrics we use to evaluate the algorithms.
	
	\begin{enumerate}
		\item Probability of a Successful Trial (PST)~\cite{tannu2019not}. This metric is used to represent the circuit output fidelity and is defined by the number of trials that give the expected result divided by the total number of trials. The expected result is obtained by executing the quantum circuit on the simulator. To precisely estimate the PST, we execute each quantum circuit on the quantum hardware for a large number of trials (8192). 
		
		\item Number of additional \texttt{CNOT} gates. This metric is related to the number of \texttt{SWAP} or \texttt{Bridge} gates inserted. This metric can show the ability of the algorithm to reduce the number of additional gates.
		
		\item Trial Reduction Factor (TRF). This metric is introduced in~\cite{das2019case} to evaluate the improvement of the throughput thanks to the multi-programming mechanism. It is defined as the ratio of the number of trials/shots needed when quantum circuits are executed independently to the number of trials/shots needed when they are executed simultaneously. 
	\end{enumerate}
	
	\subsubsection{Comparison}
	Several published qubit mapping algorithms~\cite{li2019tackling,wille2019mapping,murali2019noise,niu2020hardware,guerreschi2018two,itoko2020optimization} and multi-programming mapping algorithms~\cite{das2019case,liu2021qucloud} are available. We choose HA~\cite{niu2020hardware} as the baseline for independent execution, a qubit mapping algorithm taking hardware topology and calibration data into consideration to achieve high circuit fidelity with a reduced number of additional gates. Due to the different hardware access and code unavailability of the state-of-the-art multi-programming algorithms, we only compare our QuMC with independent executions to show the impact of the multi-programming mechanism. Moreover, our qubit partition algorithms can also be applied to the qubit mapping algorithm for independent executions if running a program on a relatively large quantum device.
	
	To summarize, the following comparisons are performed: 
	\begin{itemize}
		\item For independent executions, we compare the partition + improved mapping transition algorithm based on HA (labeled as PHA) versus HA to show the impact of partition on large quantum hardware for a small circuit. 
		
		\item For simultaneous executions, we compare our QuMC framework, 1) GSP + improved mapping transition (labeled as GSP) and 2) QHSP + improved mapping transition (labeled as QHSP), with independent executions, HA and PHA, to report the fidelity loss due to simultaneous executions of multiple circuits.
	\end{itemize}
	A detailed summary of the comparisons for independent and simultaneous executions is shown in Table~\ref{tab:methods}. Note that, PHA allows each quantum circuit to be executed on the best partition selected according to the partition fidelity score metric.

	\begin{table}[!htp]
		\caption{\label{tab:methods}A summary of comparisons for independent and simultaneous executions.} 
		\centering
		\begin{threeparttable}
			\begin{tabular}{|c|c|c|c|c|}
				\toprule        
				Comparison & \multicolumn{2}{c|}{Independent} & \multicolumn{2}{c|}{Simultaneous}     \cr
				\hline		                   
				Methods & HA & PHA & GSP & QHSP \cr
				\hline
				Partition & N/A & Algorithm.~\ref{algo2} & Algorithm.~\ref{algo1} & Algorithm.~\ref{algo2} \cr
				\hline
				Mapping & \cite{niu2020hardware} & \multicolumn{3}{c|}{Algorithm.~\ref{algo3}} \cr
				
				\bottomrule
			\end{tabular}
			\begin{tablenotes}
				\item 	\footnotesize{HA method does not include partition process.}
			\end{tablenotes}
			
		\end{threeparttable}
		
	\end{table}
	

	\begin{table}[!htp]
		\caption {\label{fig:benchmarks}Information of benchmarks.} 
		\begin{center}

			\includegraphics[scale=0.8]{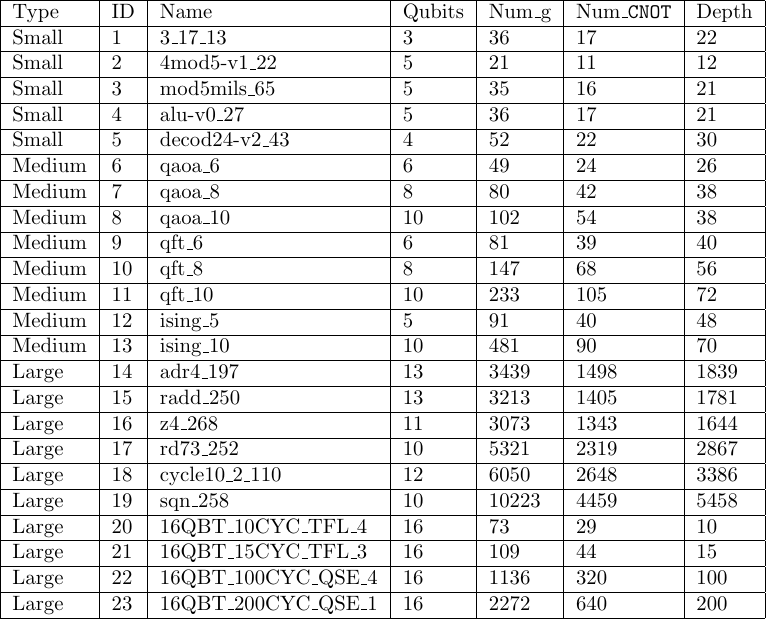}
			\centering
			\begin{tablenotes}
				\footnotesize
				\item Qubits: number of qubits. Num\_g: number of gates. Num\_\texttt{CNOT}: number of  \texttt{CNOTs}. Depth: circuit depth.

			\end{tablenotes}
		\end{center}
	\end{table}
	
	\subsubsection{Benchmarks}
	We evaluate our QuMC framework by executing a list of different-size benchmarks at the same time on two quantum devices, IBM Q 27 Toronto (ibmq\_toronto) and IBM Q 65 Manhattan (ibmq\_manhattan). 
	The benchmarks are collected from QUEKO circuits~\cite{tan2020optimality}, application-specific benchmarks, and RevLib~\cite{WGT+:2008}. These benchmarks are widely used in the quantum community and their details are shown in Table~\ref{fig:benchmarks}. We execute small quantum circuits with shallow-depth on the selected two quantum devices since only they can obtain reliable results. For medium and large quantum circuits, we compile them on the chips without hardware execution.
	
	\subsubsection{Algorithm configurations}
	Here, we consider the algorithm configurations of different multi-programming and standalone mapping approaches. We select the best initial mapping out of ten attempts for HA, PHA, GSP, and QHSP.
	Weight parameter $W$ in the cost function (see \eqref{eq:5}) is set to 0.5 and the size of the extended layer is set to 20. Parameters $\alpha_1$ and $\alpha_2$ are set to 0.5 respectively to consider equally the swap distance and swap error rate. 
	
	For the experiments of simultaneous executions of multiple different-size circuits (Section~\ref{multiple_circuits}), the weight parameter $\lambda$ of QHSP (see \eqref{eq:2}) is set to $2$ because of the relatively large number of \texttt{CNOT} gates in benchmarks, whereas for the deuteron experiment (Section~\ref{deuteron}), $\lambda$ is set to $1$ because of the small number of \texttt{CNOTs} of the parameterized circuit. The threshold $\delta$ for post qubit partition is set to 0.1 to ensure the multi-programming reliability. Due to the expensive cost of SRB, we perform SRB only on IBM Q 27 Toronto and collect the pairs with significant crosstalk effect. Only the collected pairs are characterized and their crosstalk properties are provided to the partition process. The experimental results on IBM Q 65 Manhattan do not consider the crosstalk effect. For each algorithm, we only evaluate the mapping transition process, which means no optimisation methods like gate commutation or cancellation are applied.
	
	The algorithm is implemented in Python and evaluated on a PC with 1 Intel i5-5300U CPU and 8 GB memory. Operating System is Ubuntu 18.04. All the experiments were performed on the IBM quantum information science kit (Qiskit)~\cite{Qiskit} and the version used is 0.21.0.
	
	\subsection{Application: simultaneous executions of multiple circuits of different sizes}
	\label{multiple_circuits}
	\subsubsection{Experimental results}
	
	\begin{figure}[!htp]
		\centering
		\begin{subfigure}{0.45\columnwidth}
			\centering
			\includegraphics[scale=0.5]{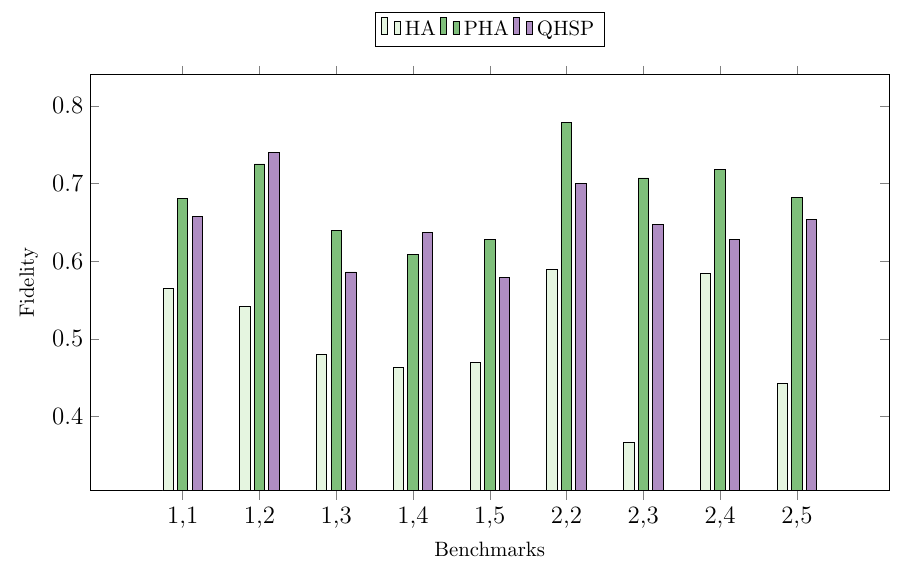}
			\label{fig:toronto_fidelity}
			\caption{}
		\end{subfigure}
		\hfill
		\begin{subfigure}{0.45\columnwidth}
			\centering
			\includegraphics[scale=0.5]{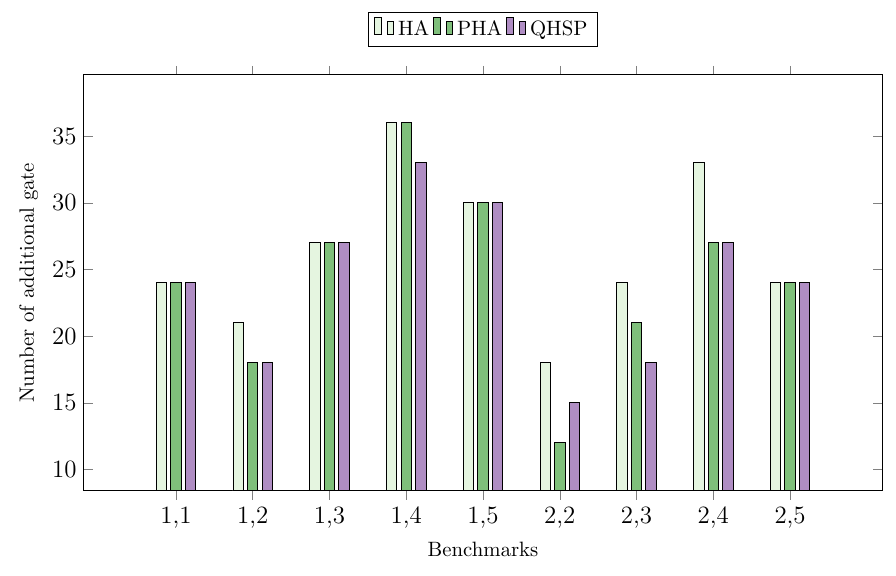}
			\label{fig:toronto_gate}
			\caption{}
		\end{subfigure}

		\caption{Comparison of average fidelity and total number of additional gates on IBM Q 27 Toronto when executing two small circuits independently and simultaneously. TRF=2. (a) Fidelity. (b) Number of additional gates.}
		\label{fig:result_toronto_figure}
	\end{figure}
	

	
		
		
	

\begin{figure}[!htp]
\centering	
\begin{subfigure}{0.45\columnwidth}
	\centering
	\includegraphics[scale=0.6]{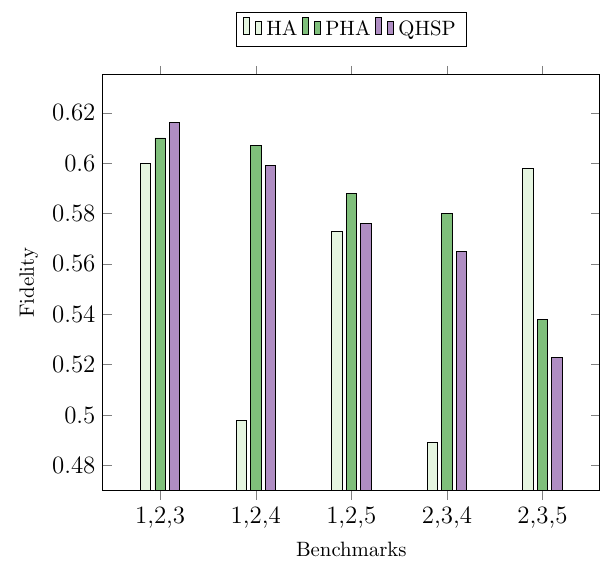}
	\label{fig:manhattan_fidelity_3}
	\caption{}
\end{subfigure}
\hfill
\begin{subfigure}{0.45\columnwidth}
	\centering
	\includegraphics[scale=0.6]{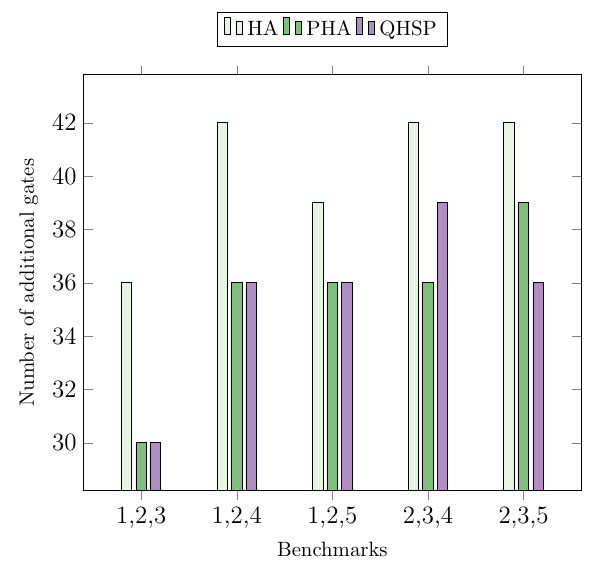}
	\label{fig:manhattan_gate_3}
	\caption{}
\end{subfigure}
\caption{Comparison of average fidelity and total number of additional gates on IBM Q 65 Manhattan when executing three small circuits independently and simultaneously. TRF=3. (a) Fidelity. (b) Number of additional gates.}
\label{fig:manhattan_result_3}
\end{figure}







\begin{figure}[!htp]

\centering
\begin{subfigure}{0.45\columnwidth}
\centering
\includegraphics[scale=0.6]{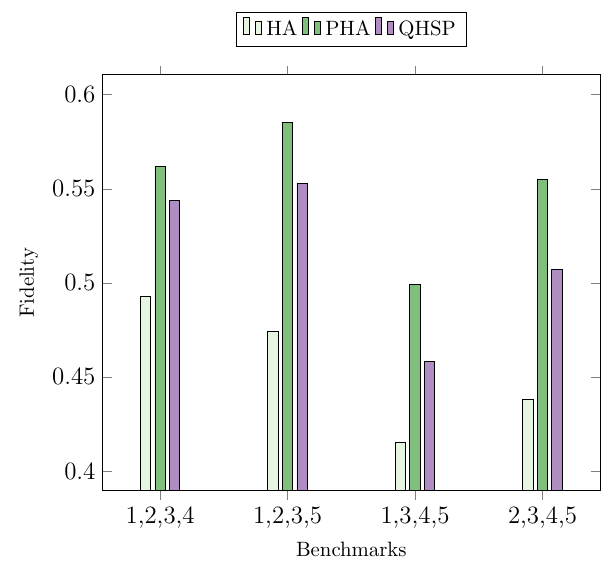}
\label{fig:manhattan_fidelity_4}
\caption{}
\end{subfigure}
\hfill
\begin{subfigure}{0.45\columnwidth}
\centering
\includegraphics[scale=0.6]{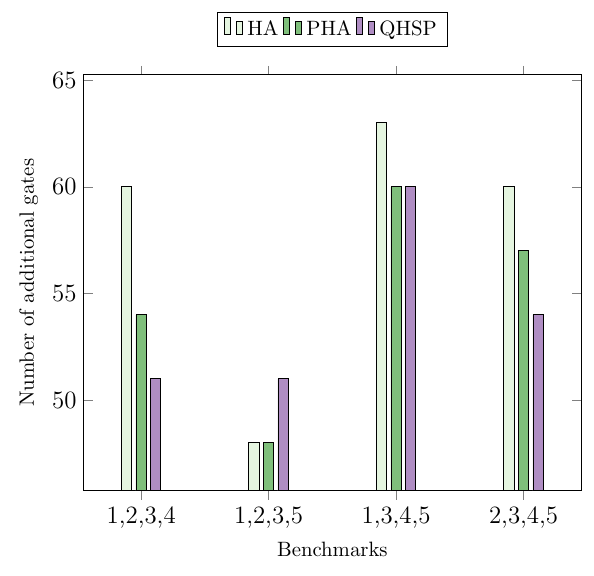}
\label{fig:manhattan_gate_4}
\caption{}
\end{subfigure}

\caption{Comparison of average fidelity and total number of additional gates on IBM Q 65 Manhattan when executing four small circuits independently and simultaneously. TRF=4. (a) Fidelity. (b) Number of additional gates.}
\label{fig:manhattan_result_4}
\end{figure}

\begin{figure}[!htp]

\centering
\begin{subfigure}{0.45\columnwidth}
\centering
\includegraphics[scale=0.6]{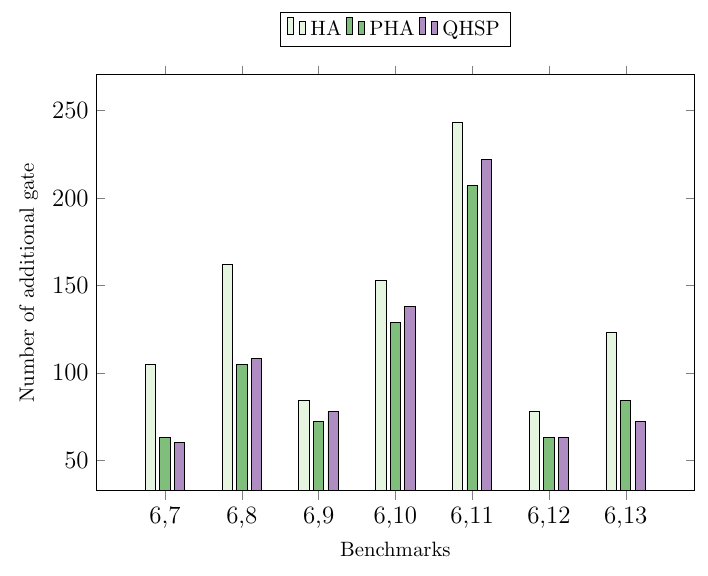}
\label{fig:toronto_large}
\caption{}
\end{subfigure}
\hfill	
\begin{subfigure}{0.45\columnwidth}
\centering
\includegraphics[scale=0.6]{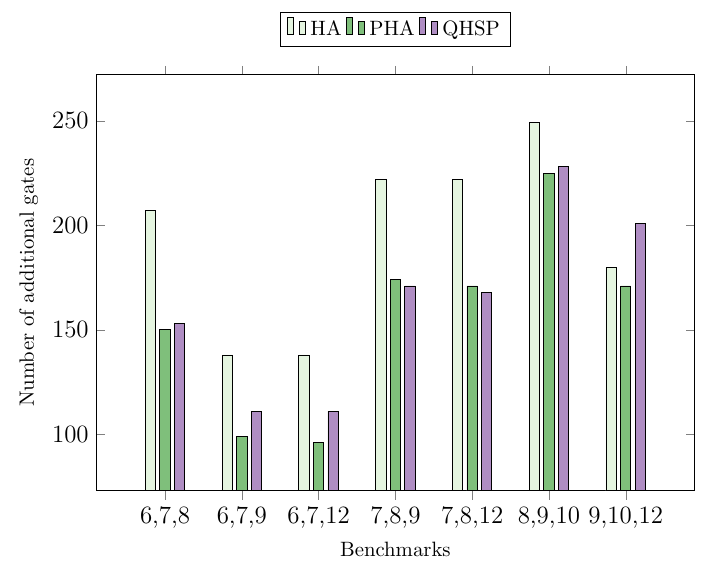}
\label{fig:manhattan_large}
\caption{}
\end{subfigure}

\caption{Comparison of total number of additional gates for medium benchmarks when (a) compiling two benchmarks on IBM Q 27 Toronto (TRF=2). (b) compiling three benchmarks on IBM Q 65 Manhattan (TRF=3).}
\label{fig:large_benchmarks}
\end{figure}

\begin{figure}[!htp]

\centering
\includegraphics[scale=0.5]{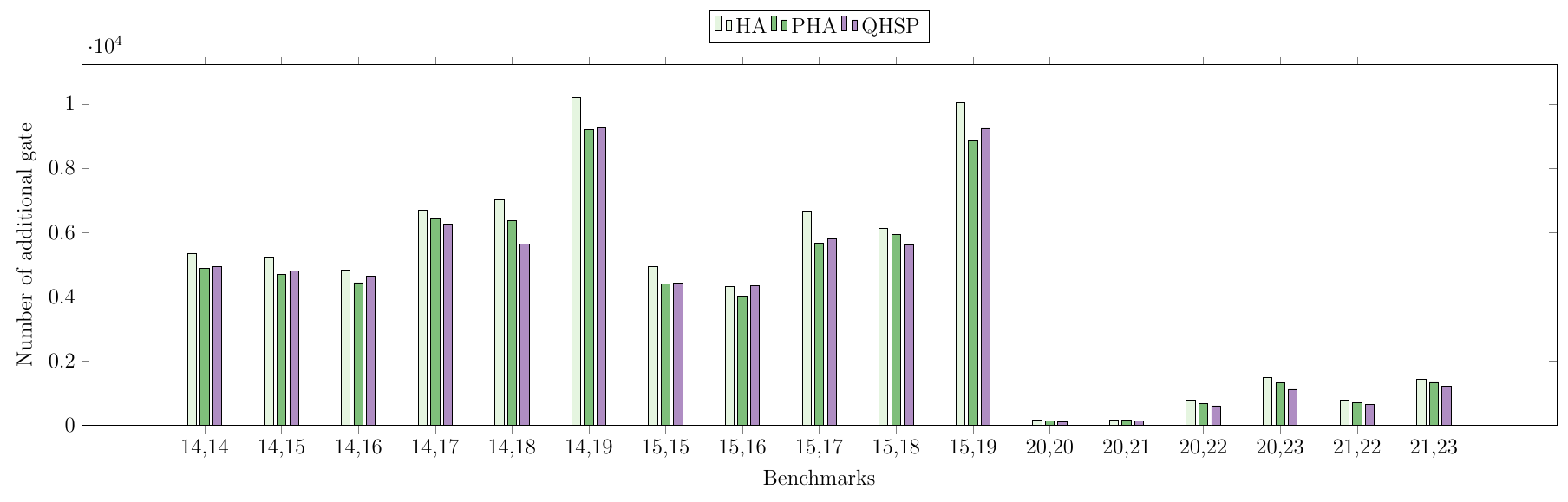}

\caption{Comparison of total number of additional gates for large benchmarks when compiling two benchmarks on IBM Q 65 Manhattan (TRF=2).}
\label{fig:large_benchmarks_2}
\end{figure}

We first run two small quantum circuits on IBM Q 27 Toronto independently and simultaneously. Results on average output state fidelity and the total number of additional gates are shown in Fig.~\ref{fig:result_toronto_figure}. Note that, all the circuit output fidelities are calculated by PST metric explained in Section~\ref{metrics}.


For independent executions, the fidelity is improved by 46.8\% and the number of additional gates is reduced by 8.7\% comparing PHA to HA. For simultaneous executions, QHSP and GSP allocate the same partitions except for the first experiment -- (ID1, ID1). In this experiment, GSP improves the fidelity by 6\% compared to QHSP. Note that partition results might be different due to the various calibration data and the choice of $\lambda$, but the difference of the partition fidelity score between the two algorithms is small. The results show that QHSP is able to allocate nearly optimal partitions while reducing runtime significantly (from exponential to polynomial complexity). Therefore, for the rest experiments, we only evaluate QHSP algorithm. Comparing QHSP (simultaneous executions) versus HA (independent executions), the fidelity is even improved by 31.8\% and the number of additional gates is reduced by 9.2\%. Whereas comparing QHSP with PHA, the fidelity is decreased by 5.4\% and the gate number is almost the same, with only 0.3\% increase. During the post-partition process, $\Delta S$ does not pass the threshold for all the combinations of benchmarks so that TRF is two, which means that the hardware throughput is improved by two times.


Next, we execute on IBM Q 65 Manhattan three and four simultaneous quantum circuits and compare the results with the independent executions. Fig.~\ref{fig:manhattan_result_3} and Fig.~\ref{fig:manhattan_result_4} show the comparison of fidelity and the number of additional gates. PHA outperforms HA for independent executions in most of the cases. Comparing QHSP with HA, the fidelity is improved by 5.3\% and 13.3\% for three and four simultaneous executions, and the inserted gate number is always reduced. Whereas the fidelities decrease by 1.5\% and 6.4\% respectively for the two cases when comparing QHSP versus PHA, and the additional gate number is always almost the same.  The threshold is still not passed for each experiment and TRF becomes three and four. 



Then, to evaluate the hardware limitations of executing multiple circuits in parallel, we set the threshold $\delta$ to 0.2. All the five small benchmarks are able to be executed simultaneously on IBM Q 65 Manhattan. Partition fidelity difference is 0.18. The average fidelity of simultaneous executions (QHSP) and independent executions (PHA) is 0.493 and 0.54, respectively, corresponding to a fidelity loss of 9.5\%.

Finally, to illustrate our QHSP algorithm's performance on medium and large benchmarks, we compile two medium-size circuits on IBM Q 27 Toronto, two medium-size circuits and three large-size circuits on IBM Q 65 Manhattan, simultaneously. We compare the results with HA and PHA for independent compilation.
Since these benchmarks are not able to obtain meaningful results due to the noise, we do not execute them on the real hardware and only use the number of additional gates as the comparison metric. The results are shown in Fig.~\ref{fig:large_benchmarks} and Fig.~\ref{fig:large_benchmarks_2}. The additional gate number is reduced by 23.2\%, 15.6\%, and 13.2\% respectively comparing QHSP with HA. When compared with PHA, the additional gate number is increased by 0.9\% and 6.4\%, and is reduced by 4.5\% respectively. All the program-wise experimental results are listed in Appendix~\ref{append}.

\subsubsection{Result analysis}

PHA is always better than HA for independent executions for two reasons: (1) The initial mapping of the two algorithms is based on a random process. During the experiment, we perform the initial mapping generation process ten times and select the best one. However, for PHA, we first limit the random process into a reliable and well-connected small partition space rather than the overall hardware space used by HA. Therefore, with only ten trials, PHA finds a better initial mapping. (2) We improve the mapping transition process of PHA, which can make a better selection between \texttt{SWAP} and \texttt{Bridge} gate. HA is shown to be sufficient for hardware with a small number of qubits, for example a 5-qubit quantum chip. If we want to map a circuit on large hardware, it is better to first limit the search space into a reliable small partition and then find the initial mapping. This qubit partition approach can be applied to general qubit mapping problems for search space limitation when large hardware is selected to map. 


Comparing simultaneous process QHSP to independent process HA, QHSP is able to outperform HA with higher fidelity and a reduced number of additional gates. The improvement is also due to the partition allocation and the enhancement of the mapping transition process as explained before. When comparing QHSP with PHA (where independent circuit is executed on the best partition), QHSP uses almost the same number of additional gates whereas fidelity is decreased less than 10\% if the threshold is set to 0.1. However, the hardware throughput increases by two and four times respectively for the two devices. Note that, it also corresponds to a huge reduction of total runtime of these circuits (waiting time + circuit execution time).


\subsection{Application: estimate the ground state energy of deuteron}
\label{deuteron}
In order to demonstrate the potential interest to apply the multi-programming mechanism to existing quantum algorithms, we investigate it on VQE algorithm.
To do this, we perform the same experiment as~\cite{gokhale2020optimization,dumitrescu2018cloud} on IBM Q 65 Manhattan, estimating the ground state energy of deuteron, which is the nucleus of a deuterium atom, an isotope of hydrogen.  

Deuteron can be modeled using a 2-qubit Hamiltonian spanning four Pauli strings: $ZI, IZ, XX,$ and $YY$~\cite{gokhale2020optimization,dumitrescu2018cloud}. If we use the naive measurement to calculate the state energy, one ansatz corresponds to four different measurements. Pauli operator grouping (labeled as PG) has been proposed to reduce this overhead by utilizing simultaneous measurement~\cite{gokhale2020optimization,kandala2017hardware,crawford2021efficient}. For example, the Pauli strings can be partitioned into two commuting families: \{$ZI, IZ$\} and \{$XX,YY$\} using the approach proposed in~\cite{gokhale2020optimization}. It allows one parameterized ansatz to be measured twice instead of four measurements in naive method.

We use a simplified Unitary Coupled Cluster ansatz with a single parameter and three gates, as described in~\cite{gokhale2020optimization,dumitrescu2018cloud}. We apply our QuMC method on the top of the Pauli operator grouping approach (labeled as QuMCPG) to estimate the ground state energy of deuteron and compare the results with PG. 

\begin{figure}[!htp]
\centering
\begin{subfigure}{0.45\columnwidth}
\centering
\includegraphics[scale=0.35]{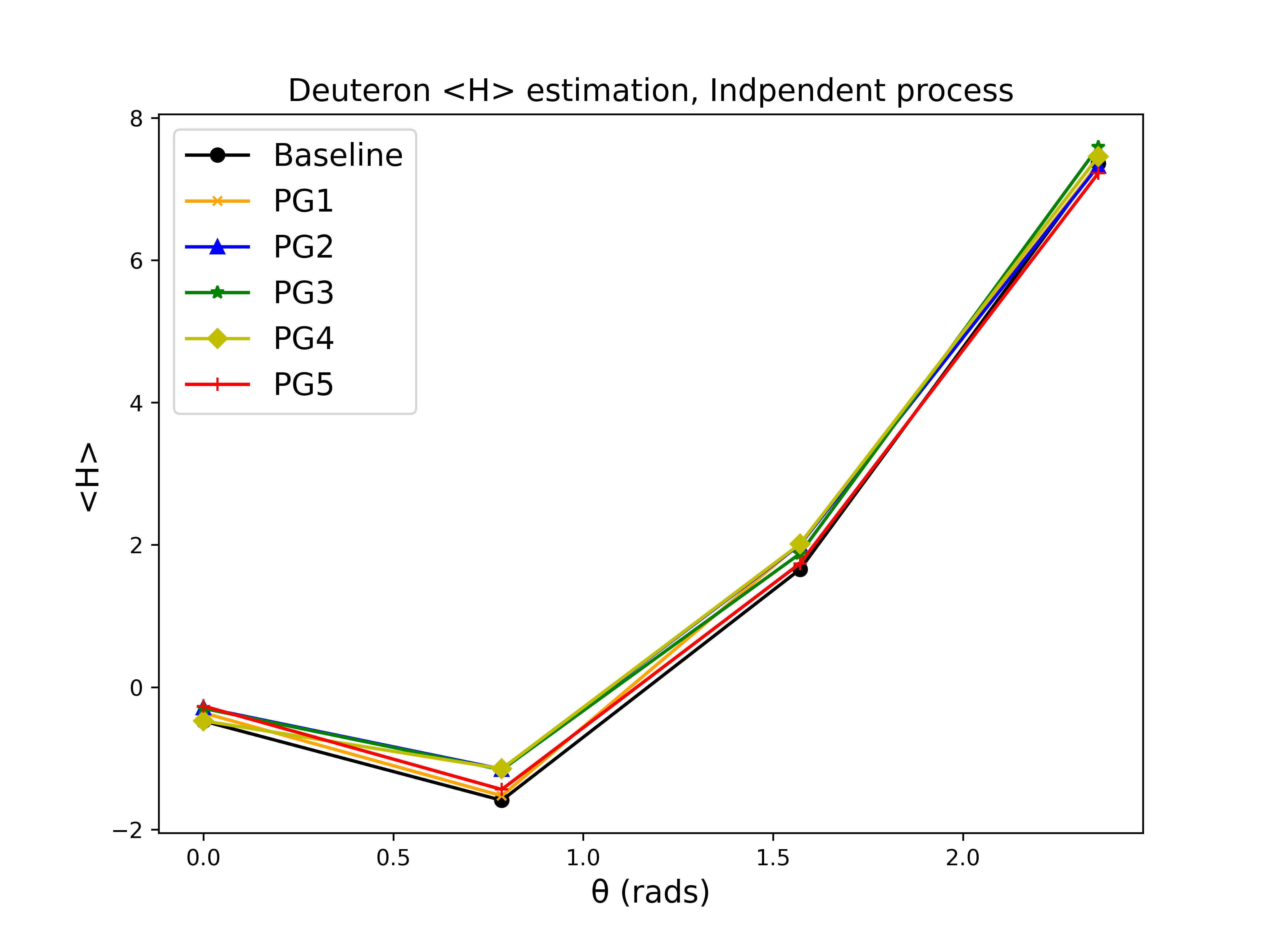}
\label{:deuteron1}
\caption{}
\end{subfigure}
\hfill
\begin{subfigure}{0.45\columnwidth}
\centering
\includegraphics[scale=0.35]{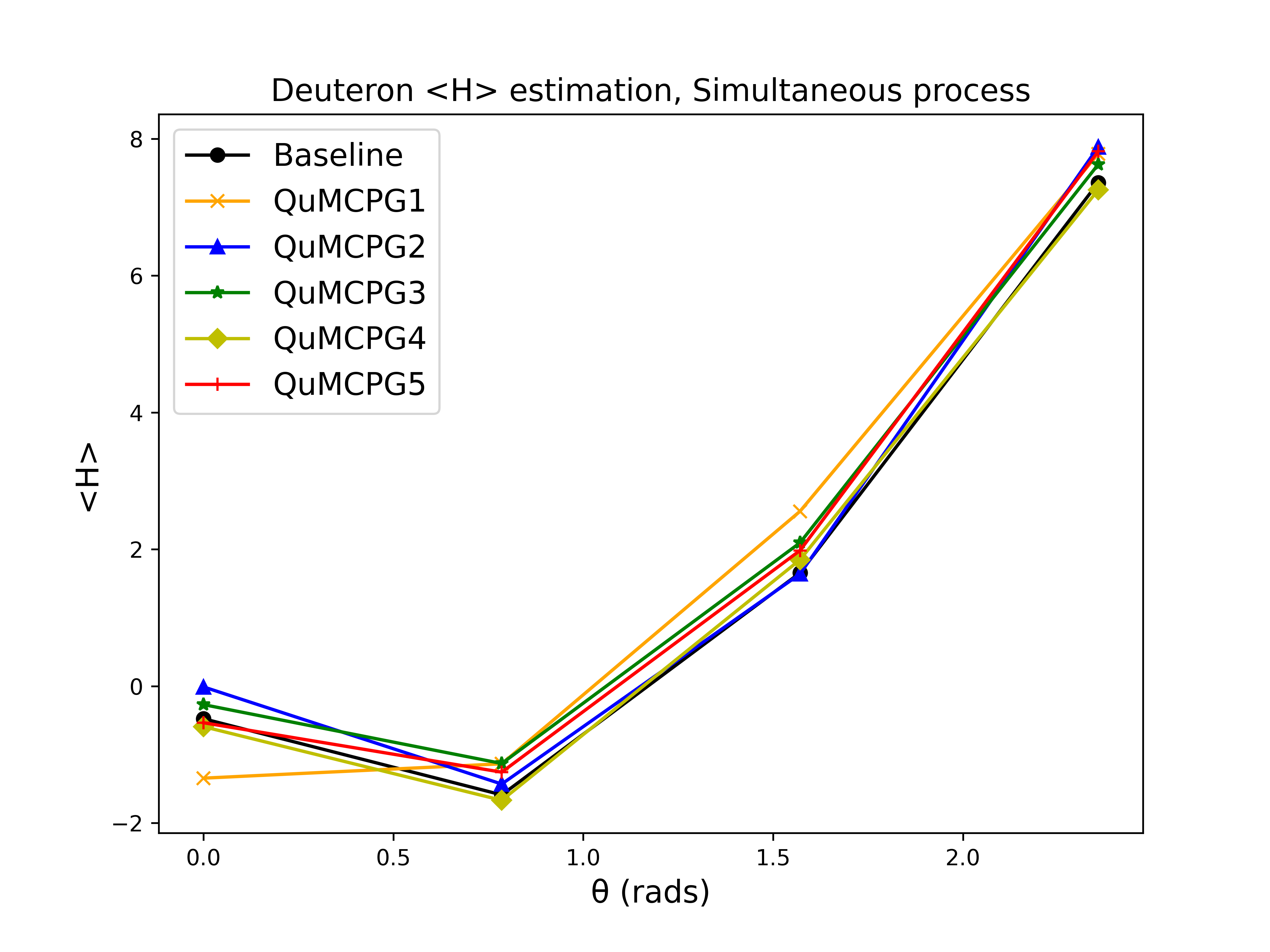}
\label{fig:deuteron2}
\caption{}
\end{subfigure}

\caption{The estimation of the ground state energy of deuteron under PG and QuMCPG with four optimisations. (a) PG result (independent process) with eight measurements. (b) QuMCPG result (simultaneous process) with one measurement. TRF=8.}
\label{fig:deuteron}
\end{figure}

In our QuMC method, the parallelism manager works with the hardware-aware multi-programming compiler to determine the number of circuits for simultaneous execution. Eight circuits are selected in order not to pass the fidelity threshold, which correspond to four parameterized circuits with four different parameters since one parameterized circuit requires two measurement circuits using PG. It is also equivalent to perform four times of optimizations. These circuits can be executed simultaneously using QuMCPG, which reduces the total circuit runtime by eight times compared with PG for independent execution. We perform this experiment five times across days with different calibration data. Note that, if we use the naive measurement, the number of measurement circuits needed will be reduced by a factor of 16. The results of the five experiments using PG (independent process) and QuMCPG (simultaneous process) are shown in Fig.~\ref{fig:deuteron}. We use simulator to perform the same experiment and set the result as baseline. The sum of the difference between the obtained result (independent or simultaneous process) and baseline (using simulator) is represented by the error rate. All the partition fidelity differences $\Delta S$ of the five experiments (on average $\Delta S$=0.06) are less than the threshold $\delta$ (set to 0.1). Compared to the baseline, the average error rates are 9\% and 13.3\% for PG and QuMCPG, respectively. Despite the augmented errors, the hardware throughput is improved by eight times. Note that, the users can tune the threshold $\delta$ according to the tolerance of the increase of error rate while using multi-programming. More information about the experimental results can be found in Table~\ref{tab:deuteron_result}. 

\begin{table}[!htp]
\centering
\begin{threeparttable}[b]
\caption{\label{tab:deuteron_result}The information of the five experiments. }

\begin{tabular}{c c c c }
\toprule                              
Experiments & $n_c$ \tnote{1} & Error rate(\%) & Hardware throughput\cr
\midrule
PG & 1 & 9 & 0.03 \\
QuMCPG & 8 & 13.3 & 0.25 \\
\bottomrule
\end{tabular}%

\begin{tablenotes}
\item [1] \small{the number of simultaneous circuit number}.
\end{tablenotes}
\end{threeparttable}

\end{table}

\section{Discussion}
\subsection{Multi-programming mechanism and fidelity loss}

\begin{figure}
\centering
\includegraphics[scale=0.6]{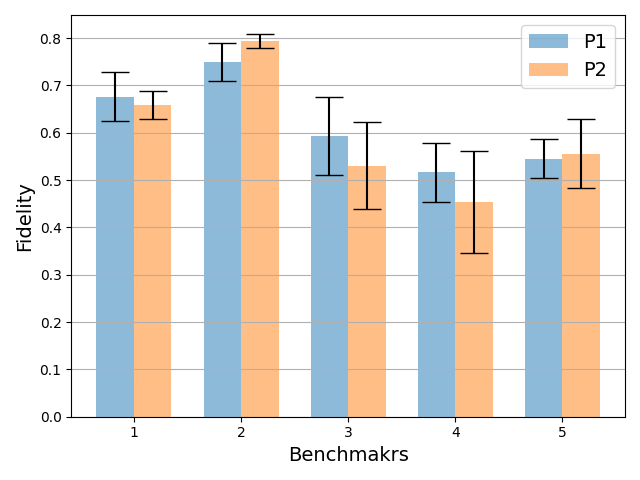}	

\caption{Comparison of fidelity on IBM Q 127 Washington when executing two of the same circuits on partitions P1 and P2 simultaneously. }
\label{fig:res-127}
\end{figure}

The aforementioned experimental results have shown that, the multi-programming mechanism can improve the hardware utilization and reduce the total circuit runtime, but with a cost of slightly losing circuit fidelity. However, the multi-programming mechanism is not always detrimental to circuit fidelity. Especially for large quantum hardware, the partition that has high fidelity is not limited to one region. We choose the largest superconducting quantum hardware, IBM 127 Q Washington (ibmq\_washington) to demonstrate it. We pick the two partitions with the highest fidelities according to our partition score metric (\ref{eq:3}) and execute two of the same circuits on the two partitions simultaneously. The score difference between the two partitions is around 0.01 and they are not adjacent to each other, so that no additional crosstalk. The benchmarks are taken from Table~\ref{fig:benchmarks} and represented by their IDs. We repeat this experiment five times and the results are shown in Fig.~\ref{fig:res-127}. P1 is the partition with the highest score and P2 with the second highest score. From the experimental results, the fidelity of the circuit on P1 cannot always outperform P2 (see benchmarks 2 and 5). It might be due to the following reasons:
(1) The calibration data are not 100\% precise. Since the partitions have almost the
same fidelity scores, the circuits executed on the two partitions should also have
similar results.
(2) The calibration data are not constant. If the circuits are waiting for a long time in
the queue, the calibration data might get updated, so that the partition with the
highest fidelity score when submitted the circuit might not be the best one when the
circuit is executed.

\subsection{Multi-programming on circuits with varying depths}
\label{sec:depth}

\begin{figure}
\centering
\includegraphics{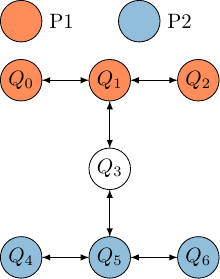}	

\caption{IBM Q 7 Nairobi hardware topology. P1 and P2 are selected to execute circuits with varying depths.}
\label{fig:7-qubit}
\end{figure}

The depths of benchmarks that we have executed on quantum hardware in Section~\ref{multiple_circuits} are not dramatically different, i.e., the longest circuit depth (decod24-v2\_43) is 2.5 times longer than the shortest one (4mod5-v1\_22). The program-wise results from Table~\ref{tab:result_toronto_fidelity}, Table~\ref{tab:result_manhattan_fidelity_3}, and Table~\ref{tab:result_manhattan_fidelity_4} have shown that the fidelity of the circuit with slightly shorter depth is not influenced by the parallel execution.

\begin{figure}
\centering
\includegraphics[scale=0.5]{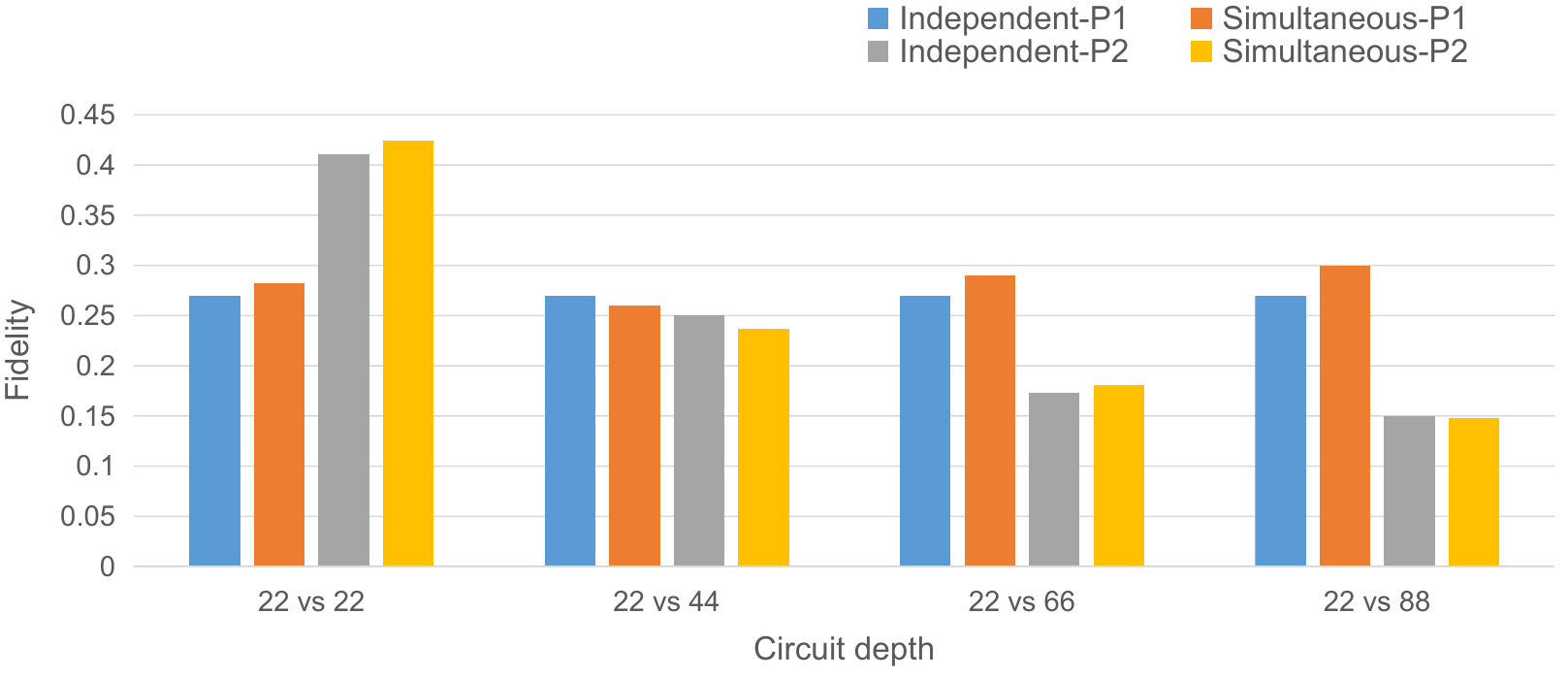}	

\caption{Comparison of fidelity on IBM Q 7 Nairobi when executing two circuits with varying depths on P1 and P2 independently and simultaneously. The blue and red columns represent the fidelities of the original circuit (fixed depth 22). The gray and yellow columns represent the fidelities of the modified circuit (depth from 22 to 88).}
\label{fig:depths}
\end{figure}

In this section, we further discuss the impact of multi-programming on circuits with comparable or dramatically different depths. We use the largest IBM public chip IBM Q 7 Nairobi (ibm\_nairobi) to perform the experiment\footnote{During the preparation of the manuscript, we do not have access to IBM private chips any more due to the end of the contract.}, and its hardware topology is shown in Fig.~\ref{fig:7-qubit}. We choose two partitions P1 and P2, which are more than one hop distance so that no additional crosstalk impact exists. First, we execute a three-qubit small circuit 3\_17\_13 with circuit depth of 22 (the information of this circuit can be found in Table~\ref{fig:benchmarks}) in P1 individually. Second, we repeat the circuit to increase its depth from one to four times and execute the circuit with varying depths in P2 individually. Finally, we execute the original circuit (depth 22) and the modified circuit (depth from 22 to 88) on P1 and P2 simultaneously. Based on the results shown in Fig.~\ref{fig:depths}, the fidelities of circuits with varying depths are not influenced by parallel executions, since all the circuit operations are scheduled “as late as possible”.

\section{Conclusion}
In this article, we presented QuMC, a multi-programming approach that allows to execute multiple circuits on a quantum chip simultaneously without losing fidelity. We introduced the parallelism manager and fidelity metric to select optimally the number of circuits to be executed at the same time. Moreover, we proposed a hardware-aware multi-programming compiler which contains two qubit partition algorithms taking hardware topology, calibration data, and crosstalk effect into account to allocate reliable partitions to different quantum circuits. We also demonstrated an improved simultaneous mapping transition algorithm which helps to transpile the circuits on quantum hardware with a reduced number of inserted gates.

We first executed a list of circuits of different sizes simultaneously and compared our algorithm with the state of the art. Experimental results showed that our QuMC can even outperform the independent executions using state of the art qubit mapping approach. Then, we investigated our QuMC approach on VQE algorithm to estimate the ground state energy of deuteron, showing the added value of applying our approach to existing quantum algorithms. The QuMC approach is evaluated on IBM hardware, but it is general enough to be adapted to other quantum hardware.


Based on the experimental result, we found that the main concern with multi-programming mechanism is a trade-off between output fidelity and the hardware throughput. For example, how one can decide which programs to execute simultaneously and how many of them to execute without losing fidelity. Here, we list several guidelines to help the user to utilize our QuMC approach.

\begin{itemize}
\item Check the target hardware topology and calibration data. The multi-programming mechanism is more suitable for a relatively large quantum chip compared to the quantum circuit and with low error rate.
\item Choose appropriate fidelity threshold for the post qubit partition process. A high threshold can improve the hardware throughput but lead to the reduction of output fidelity. It should be set carefully depending on the size of the benchmark. For benchmarks of small size that we used in experiments, it is reasonable to set the threshold to 0.1.
\item The number of circuits that can be executed simultaneously will mainly depend on the fidelity threshold and the calibration data of the hardware. 
\item The QHSP algorithm is suggested for the partition process due to efficiency and GSP is recommended to evaluate the quality of the partition algorithms. Using both algorithms, one can explore which circuits can be executed simultaneously and how many of them within the given fidelity threshold.

\end{itemize}

Quantum hardware development with more and more qubits will enable execution of multiple quantum programs simultaneously and possibly a linchpin for quantum algorithms requiring parallel sub-problem executions. The Variational Quantum Algorithm is becoming a leading strategy to demonstrate quantum advantages for practical applications. In such algorithms, the preparation of parameterized quantum state and the measurement of expectation value are realized on shallow circuits~\cite{zhang2020shallow}. Taking VQE as an example, the Hamiltonian can be decomposed into several Pauli operators and simultaneous measurement by grouping Pauli operators have been proposed in~\cite{gokhale2020optimization,kandala2017hardware,crawford2021efficient} to reduce the overhead of the algorithm. Based on our experiment, we have shown that the overhead of VQE can be further improved by executing several sets of Pauli operators simultaneously using a multi-programming mechanism. 
For future work, we would like to apply our QuMC to other variational quantum algorithms such as VQLS or VQC to prepare states in parallel and reduce the overhead of these algorithms. Moreover, in our qubit partition algorithms, we take the crosstalk effects into consideration by characterizing them and adding them to the fidelity score of the partition, which is able to avoid the crosstalk error in a high level. There are some other approaches of eliminating the crosstalk error, for example inserting barriers between simultaneous \texttt{CNOTs} to avoid crosstalk in a gate-level~\cite{murali2020software}. 
However, it has some challenges of trading-off between crosstalk and decoherence. More interesting tricks for crosstalk mitigation need to be targeted for simultaneous executions. 

\section*{Supplementary material}
The source code of the algorithms used in this paper is available
on the Github repository~\url{https://github.com/peachnuts/Multiprogramming}.

\section*{Acknowledgment}
This work is funded by the QuantUM Initiative of the Region Occitanie, University of Montpellier and IBM Montpellier. The authors are very grateful to Adrien Suau for the helpful suggestions and feedback on an early version of this manuscript. We acknowledge use of the IBM Q for this work. The views expressed are those of the authors and do not reflect the official policy or position of IBM or the IBM Q team.

\bibliographystyle{plainnat}
\bibliography{bibliography}

\begin{thebibliography}{42}
\providecommand{\natexlab}[1]{#1}
\providecommand{\url}[1]{\texttt{#1}}
\expandafter\ifx\csname urlstyle\endcsname\relax
  \providecommand{\doi}[1]{doi: #1}\else
  \providecommand{\doi}{doi: \begingroup \urlstyle{rm}\Url}\fi

\bibitem[Ash-Saki et~al.(2020{\natexlab{a}})Ash-Saki, Alam, and
  Ghosh]{ash2020analysis}
Abdullah Ash-Saki, Mahabubul Alam, and Swaroop Ghosh.
\newblock Analysis of crosstalk in nisq devices and security implications in
  multi-programming regime.
\newblock In \emph{Proceedings of the ACM/IEEE International Symposium on Low
  Power Electronics and Design}, pages 25--30, 2020{\natexlab{a}}.
\newblock \doi{https://doi.org/10.1145/3370748.3406570}.

\bibitem[Ash-Saki et~al.(2020{\natexlab{b}})Ash-Saki, Alam, and
  Ghosh]{ash2020experimental}
Abdullah Ash-Saki, Mahabubul Alam, and Swaroop Ghosh.
\newblock Experimental characterization, modeling, and analysis of crosstalk in
  a quantum computer.
\newblock \emph{IEEE Transactions on Quantum Engineering}, 2020{\natexlab{b}}.
\newblock \doi{https://doi.org/10.1109/TQE.2020.3023338}.

\bibitem[Bialczak et~al.(2010)Bialczak, Ansmann, Hofheinz, Lucero, Neeley,
  O’Connell, Sank, Wang, Wenner, Steffen, et~al.]{bialczak2010quantum}
Radoslaw~C Bialczak, Markus Ansmann, Max Hofheinz, Erik Lucero, Matthew Neeley,
  AD~O’Connell, Daniel Sank, Haohua Wang, James Wenner, Matthias Steffen,
  et~al.
\newblock Quantum process tomography of a universal entangling gate implemented
  with josephson phase qubits.
\newblock \emph{Nature Physics}, 6\penalty0 (6):\penalty0 409--413, 2010.
\newblock \doi{https://doi.org/10.1038/nphys1639}.

\bibitem[Bravo-Prieto et~al.(2020)Bravo-Prieto, LaRose, Cerezo, Subasi, Cincio,
  and Coles]{bravo2020variational}
Carlos Bravo-Prieto, Ryan LaRose, Marco Cerezo, Yigit Subasi, Lukasz Cincio,
  and Patrick Coles.
\newblock Variational quantum linear solver: A hybrid algorithm for linear
  systems.
\newblock \emph{Bulletin of the American Physical Society}, 65, 2020.

\bibitem[Calderbank and Shor(1996)]{calderbank1996good}
A~Robert Calderbank and Peter~W Shor.
\newblock Good quantum error-correcting codes exist.
\newblock \emph{Physical Review A}, 54\penalty0 (2):\penalty0 1098, 1996.
\newblock \doi{https://doi.org/10.1103/PhysRevA.54.1098}.

\bibitem[Cerezo et~al.(2021)Cerezo, Arrasmith, Babbush, Benjamin, Endo, Fujii,
  McClean, Mitarai, Yuan, Cincio, et~al.]{cerezo2021variational}
Marco Cerezo, Andrew Arrasmith, Ryan Babbush, Simon~C Benjamin, Suguru Endo,
  Keisuke Fujii, Jarrod~R McClean, Kosuke Mitarai, Xiao Yuan, Lukasz Cincio,
  et~al.
\newblock Variational quantum algorithms.
\newblock \emph{Nature Reviews Physics}, 3\penalty0 (9):\penalty0 625--644,
  2021.
\newblock \doi{https://doi.org/10.1038/s42254-021-00348-9}.

\bibitem[Crawford et~al.(2021)Crawford, van Straaten, Wang, Parks, Campbell,
  and Brierley]{crawford2021efficient}
Ophelia Crawford, Barnaby van Straaten, Daochen Wang, Thomas Parks, Earl
  Campbell, and Stephen Brierley.
\newblock Efficient quantum measurement of pauli operators in the presence of
  finite sampling error.
\newblock \emph{Quantum}, 5:\penalty0 385, 2021.
\newblock \doi{https://doi.org/10.22331/q-2021-01-20-385}.

\bibitem[Cross et~al.(2017)Cross, Bishop, Smolin, and Gambetta]{cross2017open}
Andrew~W Cross, Lev~S Bishop, John~A Smolin, and Jay~M Gambetta.
\newblock Open quantum assembly language.
\newblock \emph{arXiv preprint arXiv:1707.03429}, 2017.

\bibitem[Cross et~al.(2019)Cross, Bishop, Sheldon, Nation, and
  Gambetta]{cross2019validating}
Andrew~W Cross, Lev~S Bishop, Sarah Sheldon, Paul~D Nation, and Jay~M Gambetta.
\newblock Validating quantum computers using randomized model circuits.
\newblock \emph{Physical Review A}, 100\penalty0 (3):\penalty0 032328, 2019.
\newblock \doi{https://doi.org/10.1103/PhysRevA.100.032328}.

\bibitem[Das et~al.(2019)Das, Tannu, Nair, and Qureshi]{das2019case}
Poulami Das, Swamit~S Tannu, Prashant~J Nair, and Moinuddin Qureshi.
\newblock A case for multi-programming quantum computers.
\newblock In \emph{Proceedings of the 52nd Annual IEEE/ACM International
  Symposium on Microarchitecture}, pages 291--303, 2019.
\newblock \doi{https://doi.org/10.1145/3352460.3358287}.

\bibitem[Dumitrescu et~al.(2018)Dumitrescu, McCaskey, Hagen, Jansen, Morris,
  Papenbrock, Pooser, Dean, and Lougovski]{dumitrescu2018cloud}
Eugene~F Dumitrescu, Alex~J McCaskey, Gaute Hagen, Gustav~R Jansen, Titus~D
  Morris, T~Papenbrock, Raphael~C Pooser, David~Jarvis Dean, and Pavel
  Lougovski.
\newblock Cloud quantum computing of an atomic nucleus.
\newblock \emph{Physical review letters}, 120\penalty0 (21):\penalty0 210501,
  2018.
\newblock \doi{https://doi.org/10.1103/PhysRevLett.120.210501}.

\bibitem[Erhard et~al.(2019)Erhard, Wallman, Postler, Meth, Stricker, Martinez,
  Schindler, Monz, Emerson, and Blatt]{erhard2019characterizing}
Alexander Erhard, Joel~J Wallman, Lukas Postler, Michael Meth, Roman Stricker,
  Esteban~A Martinez, Philipp Schindler, Thomas Monz, Joseph Emerson, and
  Rainer Blatt.
\newblock Characterizing large-scale quantum computers via cycle benchmarking.
\newblock \emph{Nature communications}, 10\penalty0 (1):\penalty0 1--7, 2019.
\newblock \doi{https://doi.org/10.1038/s41467-019-13068-7}.

\bibitem[et~al.(2019)]{Qiskit}
H{\'e}ctor~Abraham et~al.
\newblock Qiskit: An open-source framework for quantum computing.
\newblock \url{https://qiskit.org/}, 2019.

\bibitem[Gambetta et~al.(2012)Gambetta, C{\'o}rcoles, Merkel, Johnson, Smolin,
  Chow, Ryan, Rigetti, Poletto, Ohki, et~al.]{gambetta2012characterization}
Jay~M Gambetta, AD~C{\'o}rcoles, Seth~T Merkel, Blake~R Johnson, John~A Smolin,
  Jerry~M Chow, Colm~A Ryan, Chad Rigetti, S~Poletto, Thomas~A Ohki, et~al.
\newblock Characterization of addressability by simultaneous randomized
  benchmarking.
\newblock \emph{Physical review letters}, 109\penalty0 (24):\penalty0 240504,
  2012.
\newblock \doi{https://doi.org/10.1103/PhysRevLett.109.240504}.

\bibitem[Gokhale et~al.(2020)Gokhale, Angiuli, Ding, Gui, Tomesh, Suchara,
  Martonosi, and Chong]{gokhale2020optimization}
Pranav Gokhale, Olivia Angiuli, Yongshan Ding, Kaiwen Gui, Teague Tomesh,
  Martin Suchara, Margaret Martonosi, and Frederic~T Chong.
\newblock Optimization of simultaneous measurement for variational quantum
  eigensolver applications.
\newblock In \emph{2020 IEEE International Conference on Quantum Computing and
  Engineering (QCE)}, pages 379--390. IEEE, 2020.
\newblock \doi{https://doi.org/10.1109/QCE49297.2020.00054}.

\bibitem[Guerreschi and Park(2018)]{guerreschi2018two}
Gian~Giacomo Guerreschi and Jongsoo Park.
\newblock Two-step approach to scheduling quantum circuits.
\newblock \emph{Quantum Science and Technology}, 3\penalty0 (4):\penalty0
  045003, 2018.
\newblock \doi{https://doi.org/10.1088/2058-9565/aacf0b}.

\bibitem[Havl{\'\i}{\v{c}}ek et~al.(2019)Havl{\'\i}{\v{c}}ek, C{\'o}rcoles,
  Temme, Harrow, Kandala, Chow, and Gambetta]{havlivcek2019supervised}
Vojt{\v{e}}ch Havl{\'\i}{\v{c}}ek, Antonio~D C{\'o}rcoles, Kristan Temme,
  Aram~W Harrow, Abhinav Kandala, Jerry~M Chow, and Jay~M Gambetta.
\newblock Supervised learning with quantum-enhanced feature spaces.
\newblock \emph{Nature}, 567\penalty0 (7747):\penalty0 209--212, 2019.
\newblock \doi{https://doi.org/10.1038/s41586-019-0980-2}.

\bibitem[Itoko et~al.(2020)Itoko, Raymond, Imamichi, and
  Matsuo]{itoko2020optimization}
Toshinari Itoko, Rudy Raymond, Takashi Imamichi, and Atsushi Matsuo.
\newblock Optimization of quantum circuit mapping using gate transformation and
  commutation.
\newblock \emph{Integration}, 70:\penalty0 43--50, 2020.
\newblock \doi{10.1016/j.vlsi.2019.10.004}.

\bibitem[Kandala et~al.(2017)Kandala, Mezzacapo, Temme, Takita, Brink, Chow,
  and Gambetta]{kandala2017hardware}
Abhinav Kandala, Antonio Mezzacapo, Kristan Temme, Maika Takita, Markus Brink,
  Jerry~M Chow, and Jay~M Gambetta.
\newblock Hardware-efficient variational quantum eigensolver for small
  molecules and quantum magnets.
\newblock \emph{Nature}, 549\penalty0 (7671):\penalty0 242--246, 2017.
\newblock \doi{https://doi.org/10.1038/nature23879}.

\bibitem[Kerenidis and Prakash(2020)]{kerenidis2020quantum}
Iordanis Kerenidis and Anupam Prakash.
\newblock Quantum gradient descent for linear systems and least squares.
\newblock \emph{Physical Review A}, 101\penalty0 (2):\penalty0 022316, 2020.
\newblock \doi{10.1103/PhysRevA.101.022316}.

\bibitem[Lanyon et~al.(2010)Lanyon, Whitfield, Gillett, Goggin, Almeida,
  Kassal, Biamonte, Mohseni, Powell, Barbieri, et~al.]{lanyon2010towards}
Benjamin~P Lanyon, James~D Whitfield, Geoff~G Gillett, Michael~E Goggin,
  Marcelo~P Almeida, Ivan Kassal, Jacob~D Biamonte, Masoud Mohseni, Ben~J
  Powell, Marco Barbieri, et~al.
\newblock Towards quantum chemistry on a quantum computer.
\newblock \emph{Nature chemistry}, 2\penalty0 (2):\penalty0 106--111, 2010.
\newblock \doi{https://doi.org/10.1038/nchem.483}.

\bibitem[Li et~al.(2019)Li, Ding, and Xie]{li2019tackling}
Gushu Li, Yufei Ding, and Yuan Xie.
\newblock Tackling the qubit mapping problem for nisq-era quantum devices.
\newblock In \emph{Proceedings of the Twenty-Fourth International Conference on
  Architectural Support for Programming Languages and Operating Systems}, pages
  1001--1014, 2019.
\newblock \doi{10.1145/3297858.3304023}.

\bibitem[Liu and Dou(2021)]{liu2021qucloud}
Lei Liu and Xinglei Dou.
\newblock Qucloud: A new qubit mapping mechanism for multi-programming quantum
  computing in cloud environment.
\newblock In \emph{2021 IEEE International Symposium on High-Performance
  Computer Architecture (HPCA)}, pages 167--178. IEEE, 2021.
\newblock \doi{https://doi.org/10.1109/HPCA51647.2021.00024}.

\bibitem[Mundada et~al.(2019)Mundada, Zhang, Hazard, and
  Houck]{mundada2019suppression}
Pranav Mundada, Gengyan Zhang, Thomas Hazard, and Andrew Houck.
\newblock Suppression of qubit crosstalk in a tunable coupling superconducting
  circuit.
\newblock \emph{Physical Review Applied}, 12\penalty0 (5):\penalty0 054023,
  2019.
\newblock \doi{https://doi.org/10.1103/PhysRevApplied.12.054023}.

\bibitem[Murali et~al.(2019)Murali, Baker, Javadi-Abhari, Chong, and
  Martonosi]{murali2019noise}
Prakash Murali, Jonathan~M Baker, Ali Javadi-Abhari, Frederic~T Chong, and
  Margaret Martonosi.
\newblock Noise-adaptive compiler mappings for noisy intermediate-scale quantum
  computers.
\newblock In \emph{Proceedings of the Twenty-Fourth International Conference on
  Architectural Support for Programming Languages and Operating Systems}, pages
  1015--1029, 2019.
\newblock \doi{10.1145/3297858.3304075}.

\bibitem[Murali et~al.(2020)Murali, McKay, Martonosi, and
  Javadi-Abhari]{murali2020software}
Prakash Murali, David~C McKay, Margaret Martonosi, and Ali Javadi-Abhari.
\newblock Software mitigation of crosstalk on noisy intermediate-scale quantum
  computers.
\newblock In \emph{Proceedings of the Twenty-Fifth International Conference on
  Architectural Support for Programming Languages and Operating Systems}, pages
  1001--1016, 2020.
\newblock \doi{https://doi.org/10.1145/3373376.3378477}.

\bibitem[Niu and Todri-Sanial(2021)]{9516713}
Siyuan Niu and Aida Todri-Sanial.
\newblock Analyzing crosstalk error in the nisq era.
\newblock In \emph{2021 IEEE Computer Society Annual Symposium on VLSI
  (ISVLSI)}, pages 428--430, 2021.
\newblock \doi{https://doi.org/10.1109/ISVLSI51109.2021.00084}.

\bibitem[Niu et~al.(2020)Niu, Suau, Staffelbach, and
  Todri-Sanial]{niu2020hardware}
Siyuan Niu, Adrien Suau, Gabriel Staffelbach, and Aida Todri-Sanial.
\newblock A hardware-aware heuristic for the qubit mapping problem in the nisq
  era.
\newblock \emph{IEEE Transactions on Quantum Engineering}, 1:\penalty0 1--14,
  2020.
\newblock \doi{10.1109/TQE.2020.3026544}.

\bibitem[Ohkura et~al.(2021)Ohkura, Satoh, and
  Van~Meter]{ohkura2021simultaneous}
Yasuhiro Ohkura, Takahiko Satoh, and Rodney Van~Meter.
\newblock Simultaneous quantum circuits execution on current and near-future
  nisq systems.
\newblock \emph{arXiv preprint arXiv:2112.07091}, 2021.

\bibitem[Pelofske et~al.(2022)Pelofske, Hahn, and
  Djidjev]{pelofske2021parallel}
Elijah Pelofske, Georg Hahn, and Hristo~N Djidjev.
\newblock Parallel quantum annealing.
\newblock \emph{Scientific Reports}, 12\penalty0 (1):\penalty0 1--11, 2022.
\newblock \doi{https://doi.org/10.1038/s41598-022-08394-8}.

\bibitem[Peruzzo et~al.(2014)Peruzzo, McClean, Shadbolt, Yung, Zhou, Love,
  Aspuru-Guzik, and O’brien]{peruzzo2014variational}
Alberto Peruzzo, Jarrod McClean, Peter Shadbolt, Man-Hong Yung, Xiao-Qi Zhou,
  Peter~J Love, Al{\'a}n Aspuru-Guzik, and Jeremy~L O’brien.
\newblock A variational eigenvalue solver on a photonic quantum processor.
\newblock \emph{Nature communications}, 5:\penalty0 4213, 2014.
\newblock \doi{https://doi.org/10.1038/ncomms5213 (2014)}.

\bibitem[Preskill(2018)]{Preskill2018quantumcomputingin}
John Preskill.
\newblock Quantum {C}omputing in the {NISQ} era and beyond.
\newblock \emph{{Quantum}}, 2:\penalty0 79, August 2018.
\newblock ISSN 2521-327X.
\newblock \doi{10.22331/q-2018-08-06-79}.

\bibitem[Proctor et~al.(2019)Proctor, Carignan-Dugas, Rudinger, Nielsen,
  Blume-Kohout, and Young]{proctor2019direct}
Timothy~J Proctor, Arnaud Carignan-Dugas, Kenneth Rudinger, Erik Nielsen, Robin
  Blume-Kohout, and Kevin Young.
\newblock Direct randomized benchmarking for multiqubit devices.
\newblock \emph{Physical review letters}, 123\penalty0 (3):\penalty0 030503,
  2019.
\newblock \doi{https://doi.org/10.1103/PhysRevLett.123.030503}.

\bibitem[Resch et~al.(2021)Resch, Gutierrez, Huh, Bharadwaj, Eckert, Loh,
  Oskin, and Tannu]{resch2021accelerating}
Salonik Resch, Anthony Gutierrez, Joon~Suk Huh, Srikant Bharadwaj, Yasuko
  Eckert, Gabriel Loh, Mark Oskin, and Swamit Tannu.
\newblock Accelerating variational quantum algorithms using circuit
  concurrency.
\newblock \emph{arXiv preprint arXiv:2109.01714}, 2021.

\bibitem[Sarovar et~al.(2020)Sarovar, Proctor, Rudinger, Young, Nielsen, and
  Blume-Kohout]{sarovar2020detecting}
Mohan Sarovar, Timothy Proctor, Kenneth Rudinger, Kevin Young, Erik Nielsen,
  and Robin Blume-Kohout.
\newblock Detecting crosstalk errors in quantum information processors.
\newblock \emph{Quantum}, 4:\penalty0 321, 2020.
\newblock \doi{https://doi.org/10.22331/q-2020-09-11-321}.

\bibitem[Shor(1997)]{doi:10.1137/S0097539795293172}
Peter~W. Shor.
\newblock Polynomial-time algorithms for prime factorization and discrete
  logarithms on a quantum computer.
\newblock \emph{SIAM Journal on Computing}, 26\penalty0 (5):\penalty0
  1484--1509, 1997.
\newblock \doi{10.1137/S0097539795293172}.

\bibitem[Tan and Cong(2021)]{tan2020optimality}
Bochen Tan and Jason Cong.
\newblock Optimality study of existing quantum computing layout synthesis
  tools.
\newblock \emph{IEEE Transactions on Computers}, 70\penalty0 (9):\penalty0
  1363--1373, 2021.
\newblock \doi{https://doi.org/10.1109/TC.2020.3009140}.

\bibitem[Tannu and Qureshi(2019)]{tannu2019not}
Swamit~S Tannu and Moinuddin~K Qureshi.
\newblock Not all qubits are created equal: a case for variability-aware
  policies for nisq-era quantum computers.
\newblock In \emph{Proceedings of the Twenty-Fourth International Conference on
  Architectural Support for Programming Languages and Operating Systems}, pages
  987--999, 2019.
\newblock \doi{https://doi.org/10.1145/3297858.3304007}.

\bibitem[Wille et~al.(2008)Wille, Gro{\ss}e, Teuber, Dueck, and
  Drechsler]{WGT+:2008}
R.~Wille, D.~Gro{\ss}e, L.~Teuber, G.~W. Dueck, and R.~Drechsler.
\newblock {RevLib}: An online resource for reversible functions and reversible
  circuits.
\newblock In \emph{{Int'l Symp. on Multi-Valued Logic}}, pages 220--225, 2008.
\newblock URL \url{http://www.revlib.org}.

\bibitem[Wille et~al.(2019)Wille, Burgholzer, and Zulehner]{wille2019mapping}
Robert Wille, Lukas Burgholzer, and Alwin Zulehner.
\newblock Mapping quantum circuits to ibm qx architectures using the minimal
  number of swap and h operations.
\newblock In \emph{2019 56th ACM/IEEE Design Automation Conference (DAC)},
  pages 1--6. IEEE, 2019.
\newblock \doi{https://doi.org/10.1145/3316781.3317859}.

\bibitem[Zhang et~al.(2021)Zhang, Gomes, Berthusen, Orth, Wang, Ho, and
  Yao]{zhang2020shallow}
Feng Zhang, Niladri Gomes, Noah~F Berthusen, Peter~P Orth, Cai-Zhuang Wang,
  Kai-Ming Ho, and Yong-Xin Yao.
\newblock Shallow-circuit variational quantum eigensolver based on
  symmetry-inspired hilbert space partitioning for quantum chemical
  calculations.
\newblock \emph{Physical Review Research}, 3\penalty0 (1):\penalty0 013039,
  2021.
\newblock \doi{https://doi.org/10.1103/PhysRevResearch.3.013039}.

\bibitem[Zhao et~al.(2020)Zhao, Xu, Lan, Chu, Tan, Yu, and Yu]{zhao2020high}
Peng Zhao, Peng Xu, Dong Lan, Ji~Chu, Xinsheng Tan, Haifeng Yu, and Yang Yu.
\newblock High-contrast z z interaction using superconducting qubits with
  opposite-sign anharmonicity.
\newblock \emph{Physical Review Letters}, 125\penalty0 (20):\penalty0 200503,
  2020.
\newblock \doi{https://doi.org/10.1103/PhysRevLett.125.200503}.

\end{thebibliography}

\onecolumn
\newpage
\appendix

\section{Supplementary experimental results}
\label{append}

The program-wise experimental results of executing two small circuits simultaneously on IBM Q 27 Toronto (Table~\ref{tab:result_toronto_fidelity}, Table~\ref{tab:result_toronto_gate}), three small circuits (Table~\ref{tab:result_manhattan_fidelity_3}, Table~\ref{tab:result_manhattan_gate_3}) and four small circuits (Table~\ref{tab:result_manhattan_fidelity_4}, Table~\ref{tab:result_manhattan_gate_4}) on IBM Q 65 Manhattan, medium and large circuits on the two devices are listed (Table~\ref{table:large_toronto}, Table~\ref{table:large_manhattan}, Table~\ref{table:large_manhattan2}).

\begin {table*}[t]
\caption {Comparison of fidelity when executing two small circuits simultaneously on IBM Q 27 Toronto.} \label{tab:result_toronto_fidelity} 
\begin{center}
	\resizebox{\textwidth}{!}{%
		\begin{threeparttable}
			\centering
			\resizebox{\textwidth}{!}{%
				\includegraphics{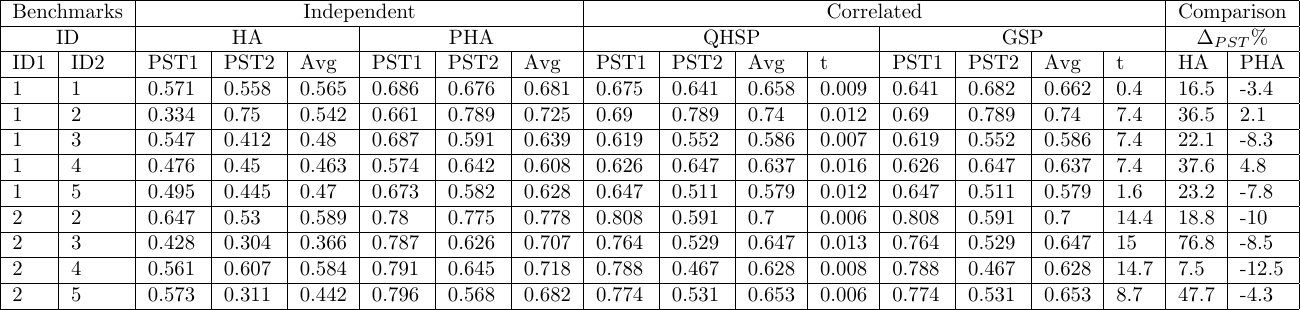}
			}
			\begin{tablenotes}
				\footnotesize
				\centering
				\item $\mathbf{Avg}$: average of PSTs. $\mathbf{t}$: runtime in seconds of the partition process. $\mathbf{\Delta_{PST}}$: comparison of average fidelity.
			\end{tablenotes}
		\end{threeparttable}
	}
\end{center}
\end{table*}

\begin {table*}
\caption {Comparison of number of additional gates when executing two small circuits simultaneously on IBM Q 27 Toronto.} \label{tab:result_toronto_gate} 
\begin{center}

\begin{threeparttable}
	\centering
	
	\includegraphics[scale=0.8]{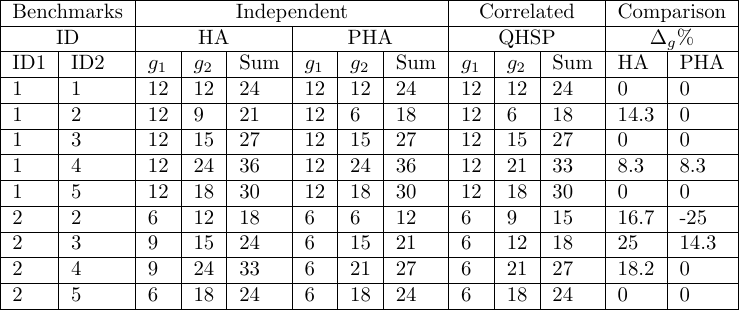}
	
	\begin{tablenotes}
		\footnotesize
		\centering
		\item $\mathbf{g}$: number of additional gates. $\mathbf{Sum}$: sum of number of additional gates. $\mathbf{\Delta_g}$: comparison of sum of number of additional gates.
	\end{tablenotes}
\end{threeparttable}

\end{center}
\end{table*}

\begin {table*}
\caption {Comparison of fidelity when executing three small circuits simultaneously on IBM Q 65 Manhattan.} \label{tab:result_manhattan_fidelity_3} 
\begin{center}
\resizebox{\textwidth}{!}{%
\begin{threeparttable}
	\centering
	\resizebox{\textwidth}{!}{%
		\includegraphics{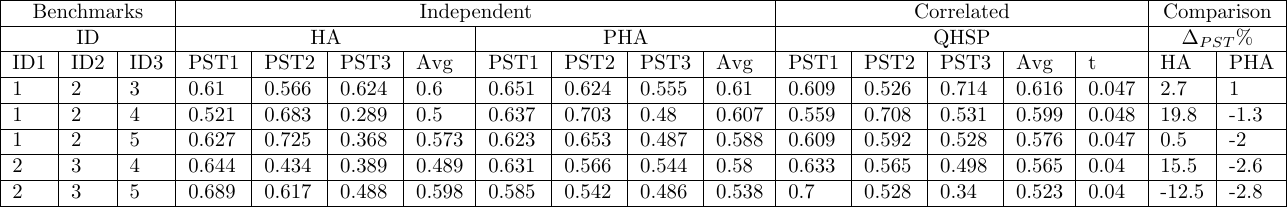}
	}
	\begin{tablenotes}
		\footnotesize
		\centering
		\item $\mathbf{Avg}$: average of PSTs. $\mathbf{t}$: runtime in seconds of the partition process. $\mathbf{\Delta_{PST}}$: comparison of average fidelity.
	\end{tablenotes}
\end{threeparttable}
}
\end{center}
\end{table*}

\begin {table*}
\caption {Comparison of number of additional gates when executing three small circuits simultaneously on IBM Q 65 Manhattan.} \label{tab:result_manhattan_gate_3} 
\begin{center}

\begin{threeparttable}
\centering

\includegraphics[scale=0.8]{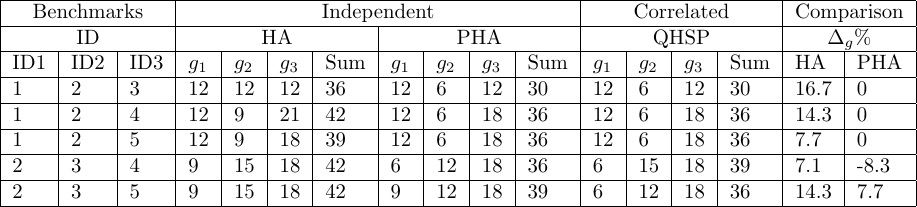}

\begin{tablenotes}
\footnotesize
\centering
\item$\mathbf{g}$: number of additional gates. $\mathbf{Sum}$: sum of number of additional gates. $\mathbf{\Delta_g}$: comparison of sum of number of additional gates.
\end{tablenotes}
\end{threeparttable}

\end{center}
\end{table*}

\begin {table*}
\caption {Comparison of fidelity when executing four small circuits simultaneously on IBM Q 65 Manhattan.} \label{tab:result_manhattan_fidelity_4} 
\begin{center}
\resizebox{\textwidth}{!}{%
\begin{threeparttable}
\centering
\resizebox{\textwidth}{!}{%
\includegraphics{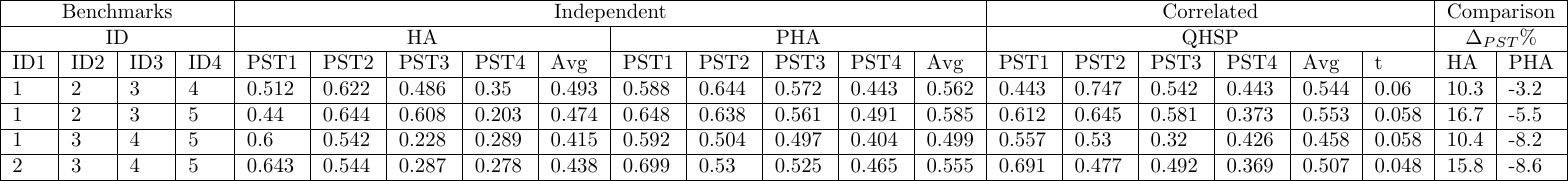}
}
\begin{tablenotes}
\footnotesize
\centering
\item $\mathbf{Avg}$: average of PSTs. $\mathbf{t}$: runtime in seconds of the partition process. $\mathbf{\Delta_{PST}}$: comparison of average fidelity.
\end{tablenotes}
\end{threeparttable}
}
\end{center}
\end{table*}

\begin {table*}
\caption {Comparison of number of additional gates when executing four small circuits simultaneously on IBM Q 65 Manhattan.} \label{tab:result_manhattan_gate_4} 
\begin{center}

\begin{threeparttable}
\centering

\includegraphics[scale=0.8]{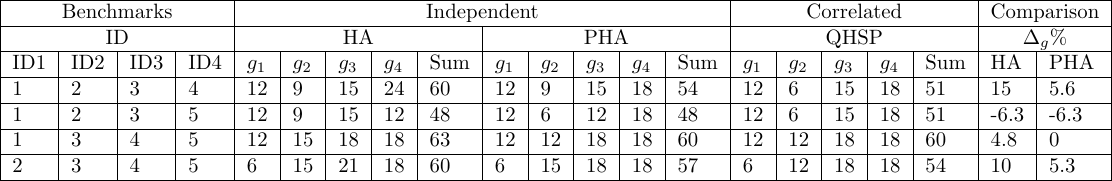}

\begin{tablenotes}
\footnotesize
\centering
\item $\mathbf{g}$: number of additional gates. $\mathbf{Sum}$: sum of number of additional gates. $\mathbf{\Delta_g}$: comparison of sum of number of additional gates.
\end{tablenotes}
\end{threeparttable}

\end{center}
\end{table*}

\begin{table}[!htp]
\caption {\label{table:large_toronto}Comparison of number of additional gates when executing two medium benchmarks on IBM Q 27 Toronto.} 

\begin{center}
\begin{threeparttable}
\centering
\includegraphics[scale=0.8]{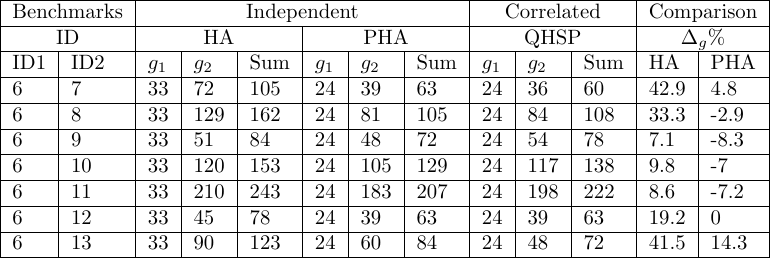}
\begin{tablenotes}
\footnotesize
\centering
\item $\mathbf{g}$: number of additional gates. $\mathbf{Sum}$: sum of number of additional gates. $\mathbf{\Delta_g}$: comparison of sum of number of additional gates.
\end{tablenotes}
\end{threeparttable}
\end{center}
\end{table}

\begin{table}[!htp]
\caption {\label{table:large_manhattan}Comparison of number of additional gates when executing three medium benchmarks on IBM Q 65 Manhattan.} 
\begin{center}
\begin{threeparttable}
\centering	

\includegraphics[scale=0.8]{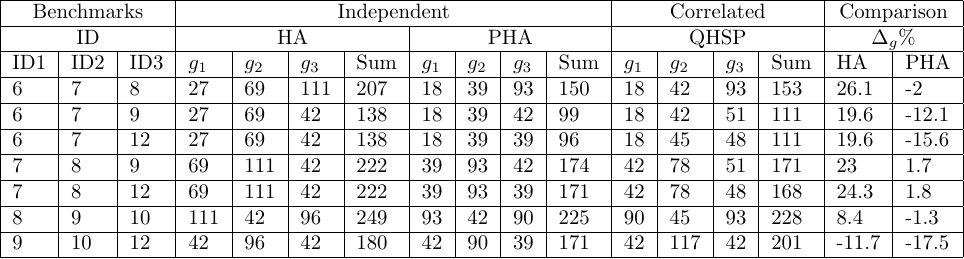}
\begin{tablenotes}
\footnotesize
\centering
\item $\mathbf{g}$: number of additional gates. $\mathbf{Sum}$: sum of number of additional gates. $\mathbf{\Delta_g}$: comparison of sum of number of additional gates.
\end{tablenotes}
\end{threeparttable}	

\end{center}
\end{table}

\begin{table}[!htp]
\caption {\label{table:large_manhattan2}Comparison of number of additional gates when executing two large benchmarks on IBM Q 65 Manhattan.} 
\begin{center}
\begin{threeparttable}
\centering	

\includegraphics[scale=0.8]{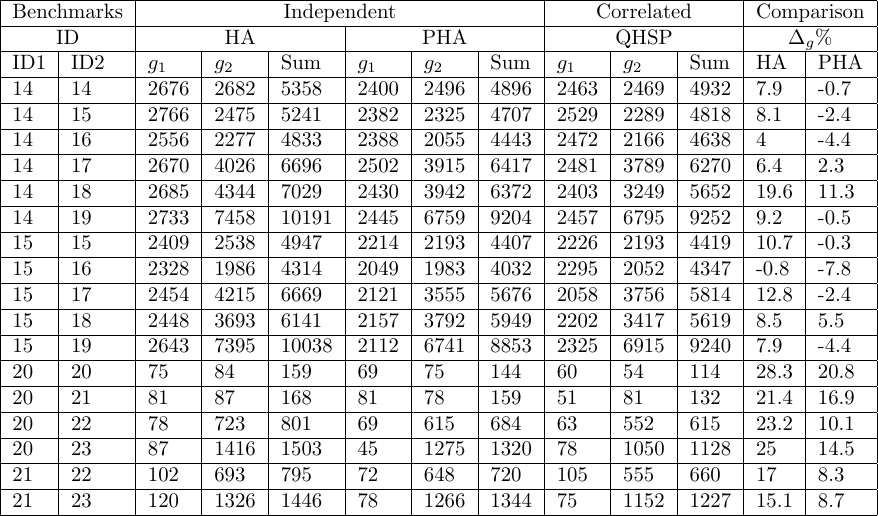}
\begin{tablenotes}
\footnotesize
\centering
\item $\mathbf{g}$: number of additional gates. $\mathbf{Sum}$: sum of number of additional gates. $\mathbf{\Delta_g}$: comparison of sum of number of additional gates.
\end{tablenotes}
\end{threeparttable}	

\end{center}
\end{table}

\end{document}